\documentclass[prl,nobibnotes,aps,reprint,a4paper,superscriptaddress,longbibliography,floatfix]{revtex4-2}
\usepackage{graphicx}   
\usepackage{epsfig}   
\usepackage{amsfonts}  
\usepackage{amsmath}  
\usepackage{amssymb}    
\usepackage{color}    
\usepackage{multirow}  
\usepackage{float}     
\usepackage[colorlinks=true,linkcolor=blue,citecolor=blue,urlcolor=blue]{hyperref}
\usepackage[utf8]{inputenc}   
\usepackage[T1]{fontenc}     
\usepackage{braket}           

\DeclareMathOperator{\Tr}{Tr}   
\newcommand{\ketbra}[2]{\ensuremath{\ket{#1}\!\bra{#2}}}

\newcommand{\brooklyn}[0]{{ibm\_\!brooklyn}}
\newcommand{\perth}[0]{{ibm\_\!perth}}
\newcommand{\tr}[1]{\textcolor{black}{#1}}
\newcommand{\lp}[1]{\textcolor{black}{#1}}
\newcommand{\lpv}[1]{\textcolor{black}{#1}}

\begin{document}

\title{Parallel tomography of \lpv{quantum non-demolition} measurements in multi-qubit devices}

\author{L.~Pereira}
\email{luciano.ivan@iff.csic.es}
\affiliation{Instituto de F\'{\i}sica Fundamental IFF-CSIC, Calle Serrano 113b, Madrid 28006, Spain}

\author{J.\thinspace J.~García-Ripoll}
\email{jj.garcia.ripoll@csic.es}
\affiliation{Instituto de F\'{\i}sica Fundamental IFF-CSIC, Calle Serrano 113b, Madrid 28006, Spain}

\author{T.~Ramos}
\email{t.ramos@csic.es}
\affiliation{Instituto de F\'{\i}sica Fundamental IFF-CSIC, Calle Serrano 113b, Madrid 28006, Spain}


\begin{abstract}
An efficient characterization of QND measurements is an important ingredient towards certifying and improving the performance and scalability of quantum processors. In this work, we introduce a parallel tomography of QND measurements that addresses single- and two-qubit readout on a multi-qubit quantum processor. We provide an experimental demonstration of the tomographic protocol on a 7-qubit IBM-Q device, characterizing the quality of conventional qubit readout as well as generalized measurements such as parity or measurement-and-reset schemes. Our protocol reconstructs the Choi matrices of the measurement processes, extracts relevant quantifiers---fidelity, QND-ness, destructiveness---and identifies sources of errors that limit the performance of the device for repeated QND measurements. We also show how to quantify measurement cross-talk and use it to certify the quality of simultaneous readout on multiple qubits.
\end{abstract}

\maketitle

\section{Introduction}

Quantum non-demolition (QND) measurements allow the repeated evaluation of an observable without changing its expected value~\cite{braginskii_quantum_1992,breuer_theory_nodate}. They have been implemented in many quantum platforms such as \lpv{atomic}~\cite{leibfried_quantum_2003,raha_optical_2020,grangier_quantum_1998,gleyzes_quantum_2007,PhysRevLett.126.253603,PhysRevLett.112.093601} \lpv{or} solid-state systems~\cite{neumann_single-shot_2010,robledo_high-fidelity_2011,nakajima_quantum_2019,xue_repetitive_2020,wallraff_approaching_2005,gomez-leon_dispersive_2022}. In superconducting quantum processors, in particular, the most widespread qubit measurement is, in its ideal form, also a QND measurement~\cite{blais_circuit_2020,vijay_observation_2011,walter_rapid_2017}. In practice, this qubit readout is not yet perfectly QND and has larger errors than single- and two-qubit gates~\cite{google_error_correction_2021,ibmq}. The origin of these measurement errors is diverse: non-dispersive interactions~\cite{boissonneault_dispersive_2009,govia_entanglement_2016}, leakage to excited states~\cite{Sank2016,wang_optimal_2021}, decoherence~\cite{Slichter2012,walter_rapid_2017}, or cross-talk~\cite{rudinger_experimental_2021,seo_measurement_2021}, and they accumulate exponentially with repeated measurements.

While state-of-the-art is adequate for restricted models of computation---e.g.~variational quantum algorithms~\cite{Cerezo2021,Bharti2022nisq} or proof-of-principle quantum error correction~\cite{krinner_realizing_2021,google_error_correction_2021,Zhao2021surfacecode}---, large-scale and fault-tolerant quantum computing schemes~\cite{calderbank_good_1996, steane_error_1996, fowler_surface_2012, campbell_roads_2017, chamberland_building_2020} require that we improve on the quality of QND measurements, through efficient, reliable, and self-consistent characterization techniques, which also help us identify and mitigate experimental errors~\cite{Eisert2020}.

Quantum tomography (QT) is a powerful and general technique to characterize the evolution of a physical system~\cite{Banaszek2013}, used e.g.~in superconducting qubits~\cite{Steffen2006,Pereira_2021_scalable,Gaikwad2022}, trapped ions~\cite{Hffner2005,Monz_2009_toffoli,Klimov_2008_optimal}, or photonic systems~\cite{Agnew_2021_tomography,Chapman_2016_experimental, Zambrano_2020_Estimation}. We proposed QND measurement tomography (QND-MT) ~\cite{pereira_qnd_tomography} as a self-consistent reconstruction of the Choi operators for a general QND detector, describing the measurement process, its dynamics, relevant quantifiers, and sources of error. A similar approach based on gate set tomography have also been recently developed~\cite{Rudinger2022midcircuit}. QND-MT is more informationally complete than a direct estimation of readout fidelity and QND-ness~\cite{Touzard2019gated,Dassonneville2020fast,blais_circuit_2020}, or a standard measurement tomography (MT) ~\cite{Lundeen2008,Fiur2001maximum, Chen2019detector,seo_measurement_2021} of the positive operator-valued measurements (POVM).

\lp{In this work, we experimentally implement an efficient parallel QND-MT to characterize the most important measurement properties of a 7-qubit IBM-Q quantum computer~\cite{ibmq}. The protocol exploits the low correlations between the qubit readout to implement a cheap parallel single-qubit characterization of each measurement, obtaining relevant quantifiers from the Choi operators such as} readout fidelity, QND-ness, and destructiveness~\cite{pereira_qnd_tomography}. We observe that the device is optimized to maximize the fidelity—--calibrated at around $\sim 98\%$ for every qubit---but not the QND-ness, \lp{which varies more accross the device and it is lower on average $\sim 96.7\%$.} QND-MT also reveals that bit flip errors are the main source of imperfections. \lp{Using a two-qubit QND-MT we quantify measurement cross-talk across device. We find a similar correlation strength between local and non-local pairs of qubits, which introduces an error of less than $1\%$ in the simultaneous execution of qubit readout. This validates the application in parallel of the single- and two-qubit tomographic protocol on the IBM-Q device, which can be executed with a constant number of circuits, avoiding the exponential scaling of a full QT. This parallelization is also extended to the post-processing of data on classical computers.}

Finally, we demonstrate the generality of QND-MT by reconstructing composite measurement processes relevant to quantum error correction protocols such as parity measurements and measurement-and-reset schemes with classical feedback. Our experiment shows that the parity measurement involves more errors---mainly non-dispersive--- than a direct QND measurement due to the presence of an entangling gate. In addition, we observe that the measurement-and-reset scheme can enhance the QND nature of the readout.

\begin{figure*}[ht!]
	\centering
	\includegraphics[width=\linewidth]{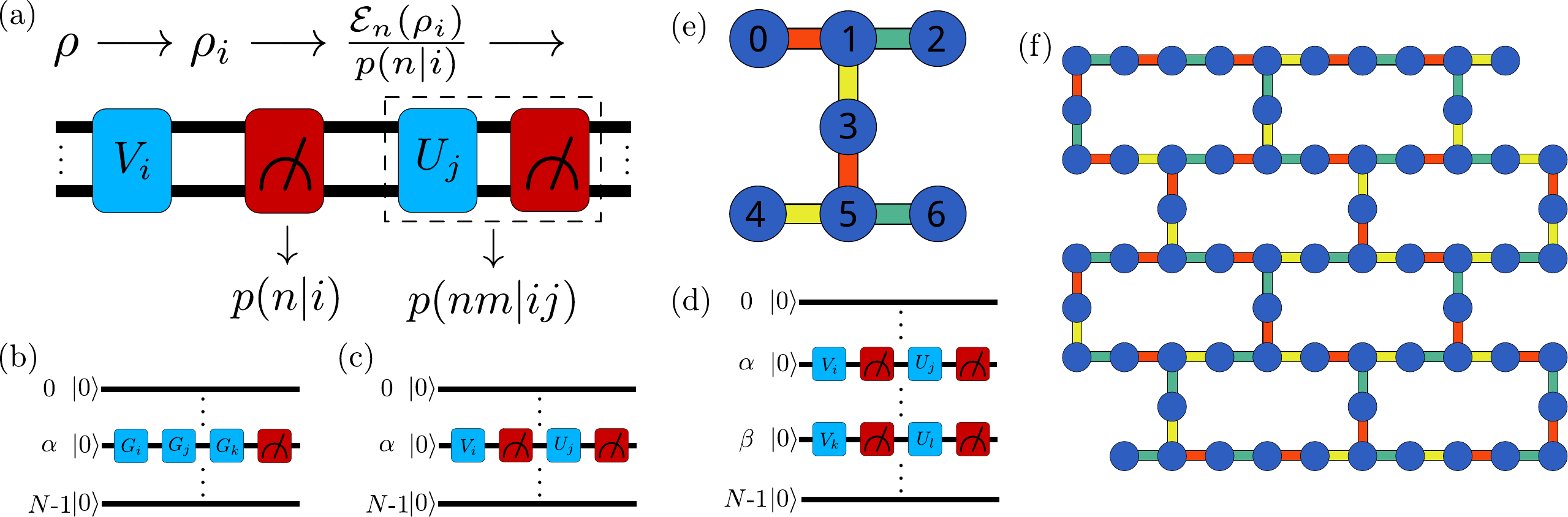}
	\caption{{\bf Quantum circuits and parallelization strategies in the QND measurement tomography.} Quantum circuits for (a) QND measurement tomography of generic measurements, (b) single-qubit gate set tomography, and (c) single-qubit and (d) two-qubit QND measurement tomography. (e)-(f) Schemes of IBM quantum devices (e) \lp{\emph{\perth}} (7 qubits) and (f) \emph{\brooklyn{}} (65 qubits). Measurements of pairs of qubits connected by a bar of the same color (red, green and yellow) are characterized simultaneously in one run of the parallel tomography.}
	\label{fig1}
\end{figure*}

\section{Results}

\subsection{QND measurement tomography on a multi-qubit device}

\lp{A generalized quantum} measurement of an $N$-qubit system in state $\rho$ is described by a set of non trace-preserving quantum processes $\mathcal{E}_n$, \lp{which add up to a trace-preserving one, $\mathcal{E}=\sum_n\mathcal{E}_n$~\cite{breuer_theory_nodate}. Each individual process determines a post-measurement state $\rho_n=\mathcal{E}_n(\rho)/p(n)$, conditioned to the measurement outcome occurring with probability $p(n)=\Tr(\mathcal{E}_n(\rho))$.} A representation of quantum processes commonly used in quantum tomography \lp{are the Choi matrices~\cite{Milz2017}. In this representation, a measurement is described by a set of Choi operators $\{\Upsilon_n\}$ whose matrix elements are given by~\cite{pereira_qnd_tomography} }
\begin{equation}
  \Upsilon_n^{ijkl}  = \braket{ij|\Upsilon_n|kl} = \braket{ i| \mathcal{E}_n( \ketbra{k}{l} ) |j},
\end{equation}
with $\{\ket{i}\}$ the basis of the measured system with dimension $d$. In terms of these matrices we can conveniently determine the dynamics of the post-measurement states $\mathcal{E}_n(\rho) = \sum_{ijkl}\Upsilon_n^{ijkl}\rho^{kl}\ket{i}\bra{j}$, the POVM elements $\Pi_n = \sum_{ijk}\Upsilon_n^{kjki}|i\rangle \langle j|$, and the measurement statistics $p(n)=\Tr\lbrace\Pi_n\rho\rbrace$, where $\rho^{kl}=\bra{k}\rho\ket{l}$ are the components of the density matrix before measurement. \lpv{Note that $\Upsilon_n$ is the transposed of the positive Choi operator $\tilde\Upsilon_n$, whose components are related by $\bra{ij}\Upsilon_n\ket{kl}=\bra{ik}\tilde\Upsilon_n\ket{jl}$~\cite{Milz2017}.}

The Choi matrices $\Upsilon_n^{ijkl}$ are a complete description of the quantum processes of a system and from them we can extract all the relevant physical properties of the measurements. We discuss three relevant quantifiers of the measurement: the readout fidelity $F$, the QND-ness $Q$, and the destructiveness $D$~\cite{pereira_qnd_tomography} (see methods). Comparing them, we can quantify the quality of the measurement for particular tasks and discriminate between different types of measurements. The readout fidelity $F$ describes the efficiency of the readout irrespective of the post-measurement state, and it is thus maximal when the POVMs are projectors, $\Pi_n=|n\rangle\langle n|$. Operationally, it is defined as the average probability of successfully detecting a state $\ket{n}$ of the computational basis, after preparing the system in the same state. The QND-ness $Q$ is the fidelity respect to an ideal measurement of an observable $O$, that is, a measurement that projects the states into the eigenvectors $|n\rangle$ of $O$ and whose Choi matrices are projectors, $\Upsilon_n=\ketbra{nn}{nn}$. QND-ness incorporates information of the post-measurement states and can be determined by the average probability that states of the computational basis $\ket{n}$ are preserved in two consecutive measurements. Finally, the destructiveness $D$ quantifies the back-action introduced by the measurement \cite{pereira_qnd_tomography}. For $D=0$, the measurement is exactly QND which means that it preserves the expected value of the observable $O$ after consecutive measurement $\braket{O}=\Tr[O\rho]=\Tr[O\mathcal{E}(\rho)]$. For $D>0$, the destructiveness signals a deviation from the QND condition, which can occur independent on how ideal the measurement is. Therefore, it is convenient to know the three quantifiers $F$, $Q$, and $D$ to provide a more complete analysis of general non-destructive measurements.

\begin{figure*}[t!]
	\centering
	\includegraphics[width=\linewidth]{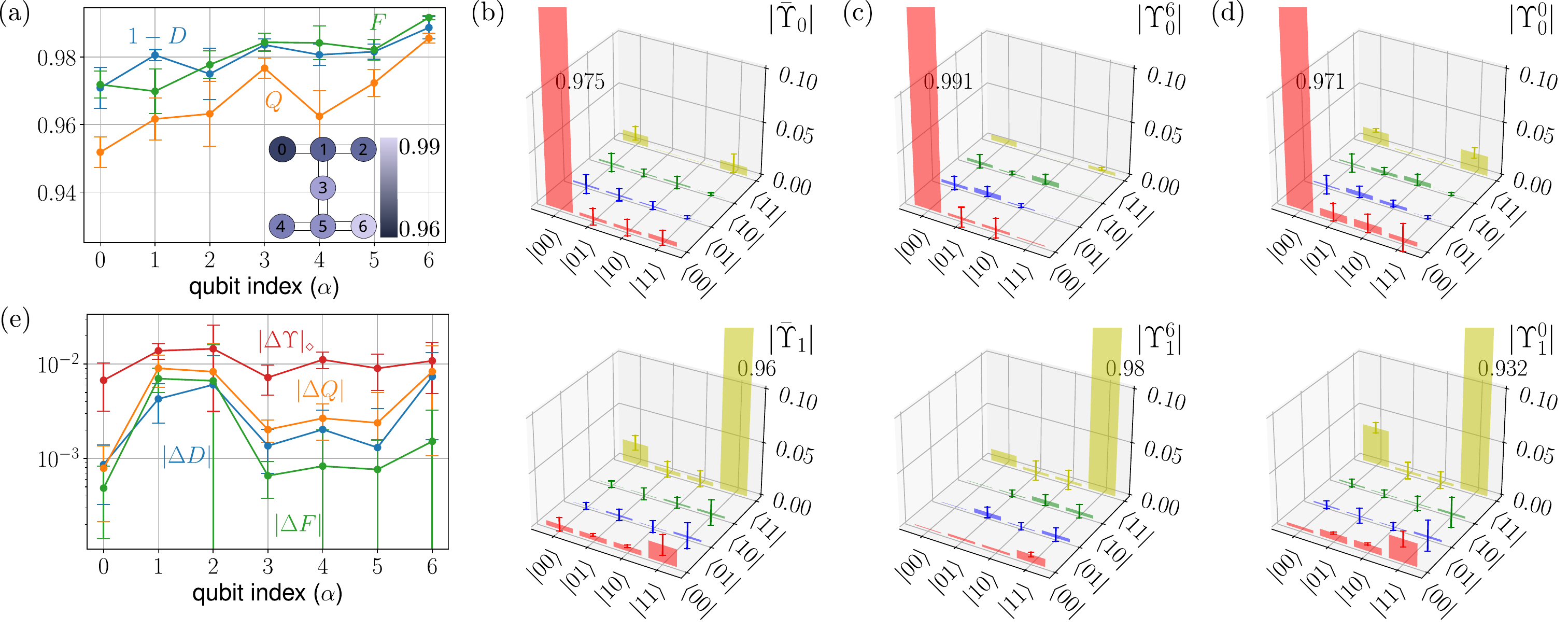}
	\caption{ {\bf Experimental tomographic characterization of single-qubit measurements \lp{in parallel}.} (a) Measurement quantifiers: Fidelity $F$ (green), QND-ness $Q$ (orange), and \lp{indestructiveness} $1-D$ (blue) for each qubit. \lp{The inset represents} average performance \lp{$(F+Q+1-D)/3$} of each qubit, as they are located on the \lp{\emph{\perth}} quantum processor. (b) Average Choi matrices $\bar{\Upsilon}_n$ of all qubits for both measurement outcomes $n=0,1$. (c)-(d) Specific Choi matrices \lp{$\Upsilon^6_n$} and \lp{$\Upsilon^0_n$} corresponding to the qubits with best and worst readout performance, respectively. \lp{(e) Error in the quantifiers and the Choi operators introduced by the parallelization of the QND-MT. Error bars are the standard deviation estimated with 5 realizations of the experiment.}}
	\label{fig2}
\end{figure*}

Our QND-MT protocol~\cite{pereira_qnd_tomography} reconstructs the Choi matrices of a QND measurement self-consistently. As shown in Fig.~\ref{fig1}(a), \lp{it consists of two applications of the measurement interspersed by} a unitary gate $V_i$ that prepares a complete basis of initial states, and a second gate $U_j$ that enables a complete set of measurements. Given sufficient statistics, this protocol provides conditional probability distributions of single QND measurements $p(n|i)$ and of consecutive measurements $p(nm|ij)$. A maximum-likelihood-based classical post-processing~\cite{Fiurasek2001maximum,James2001measurement,Shang2017superfast} \lp{transforms $p(n|i)$ and $p(nm|ij)$} into a set of physically admissible set of \lp{POVM elements $\{\Pi_n\}$ and Choi matrices $\{\Upsilon_n\}$, requiring to solve a total of ${\cal N}+1$ optimization problems, with ${\cal N}$ the number of outcomes} (see methods). The full characterization of QND detectors with $N$ qubits and \lp{${\cal N}=2^N$ possible} outcomes demands reconstructing $2^N$ Choi operators of size $4^{N}$. In the general case, using a strategy based on Pauli observables, QND-MT requires a total of $18^N$ circuits, corresponding to $6^N$ initial gates \lp{prepared with the tensor product of} $V_i\in\{ I, \sigma_x, e^{\mp i\pi\sigma_y/4}, e^{\mp i\pi\sigma_x/4}\}$ and $3^N$ intermediate unitaries \lp{given by the tensor products of} $U_i\in\{ I,  e^{-i\pi\sigma_y/4}, e^{-i\pi\sigma_x/4}\}$, with $I$ and $\sigma_j$ the identity and Pauli operators. \lp{Figures \ref{fig1}(b)-(d) show the circuits for the particular case of a single-qubit and two-qubits, as explained below.}

The exponential scaling in the number of circuits makes quantum tomography—in the QND-MT or in any other form—unfeasible for systems with large numbers of qubits. This scaling may be \tr{avoided} if the measurements of separate quantum subsystems are shown to be independent. \tr{QND measurements in superconducting circuits are implemented via dispersive readout~\cite{blais_circuit_2020, vijay_observation_2011,walter_rapid_2017}, where each qubit is coupled to an off-resonance cavity, on which one performs homodyne detection to individually extract the outcome of each qubit state. These readouts is thus built to be independent of each other, but imperfections in the device can lead to cross-talk between the qubit measurements ~\cite{seo_measurement_2021, Arute2019}. Nevertheless, these correlations can be characterized by two-qubit QND-MT of each pair of qubits. If the correlations are weak enough, it is possible to execute} a highly parallelized and scalable QND-MT for multi-qubit detectors, reducing the number of circuits and the classical post-processing time.

\subsection{\tr{Parallel single-qubit QND measurement tomography}}

\lp{Let us first discuss the tomographic reconstruction of every single-qubit measurement of a quantum processor with $N$ qubits. This means using QND-MT} to reconstruct $2N$ single-qubit Choi matrices $\Upsilon^\alpha_n$, \lp{for qubits} $\alpha=1,...,N$ and $n=0,1$. \lp{Single-qubit QND-MT applies two measurements $\Upsilon^\alpha_n$ in between single-qubit gates $V_i$ and $U_j$, as shown in Fig.\ref{fig1}(c), requiring the evaluation of $18$ different circuits. For each pair of measurement outcomes the associated Choi matrix $\Upsilon^\alpha_n$ is estimated as the solution of a maximum likelihood optimization problem.}

Initialization and gate errors \lp{are accounted for in this optimization by means of single-qubit} gate set tomography (GST)~\cite{Dehollain2016gst, Rudinger2022midcircuit, Nielsen2021gst, Rudinger2021gst2crosstalk}. GST \lp{self-consistently characterizes} the initial state, the final POVM measurement, and a complete set of linearly independent gates $G_i$ of a device. In our case, $G_i\in\{ I, \sigma_x,  e^{-i\pi\sigma_y/4}, e^{-i\pi\sigma_x/4}  \}$ \lp{are the gates used} to implement all the $V_i$ and $U_i$ operations \lp{from QND-MT. Note that GST requires $64$ circuits, each composed of three gates $G_i$ and a measurement, as shown in Fig.~\ref{fig1}(b).}

\lp{The execution of the circuits of the $N$ single-qubit QND-MTs and GST circuits can be efficiently parallelized, applying the single-qubit operations simultaneously as sketched in Figs.~\ref{fig1}(b-c). This reduces the total number of experiments from $\mathcal{O}(N)$ down to the sum of $18$ QMD-MT and $64$ GST circuits. With this refinement, we studied the readout of all qubits in the IBM quantum device \emph{\perth}.} This processor has a quantum volume \cite{quantum_volume} of 32 and CLOPS (Circuit Layer operations per second) of \tr{$2.9\times10^3$}~\cite{clops}, and a qubit connectivity graph shown in Fig.~\ref{fig1}(e). \tr{Each circuit was evaluated with $2^{13}$ shots}, resulting in an experiment that takes approximately 2 minutes, with classical post-processing of \tr{30} seconds on a Ryzen-7 5800H processor with 8 cores. 

From the reconstructed Choi matrices of each qubit, we derived the three quantifiers---readout fidelity $F$, the QND-ness $Q$, and the \lp{indestructiveness} $1-D$---shown in Figure~\ref{fig2}(a). \lp{The \emph{\perth}} processor exhibits readout fidelities between \tr{$0.969$} and \tr{$0.992$}, with an average of $\bar{F}=0.98$. QND-ness varies much more along the device, ranging from \tr{$0.951$} in the qubit $\alpha=0$ to \tr{$0.987$} in $\alpha=6$, with an average of \tr{$\bar{Q}=0.967$}. \tr{Indestructiveness behaves similarly to fidelity} except for qubit $\alpha=1$ and ranges between \tr{$0.97$} and \tr{$0.991$}. The arithmetic mean of $F$, $Q$, and $1-D$\tr{---see colormap in Fig.~\ref{fig2}(a)}---characterizes the performance across the device: qubits in the upper sector ($\alpha=0, 1, 2$) perform notably worse than those in the lower half of the chip.

\begin{figure*}[t!]
	\centering
	\includegraphics[width=0.9\linewidth]{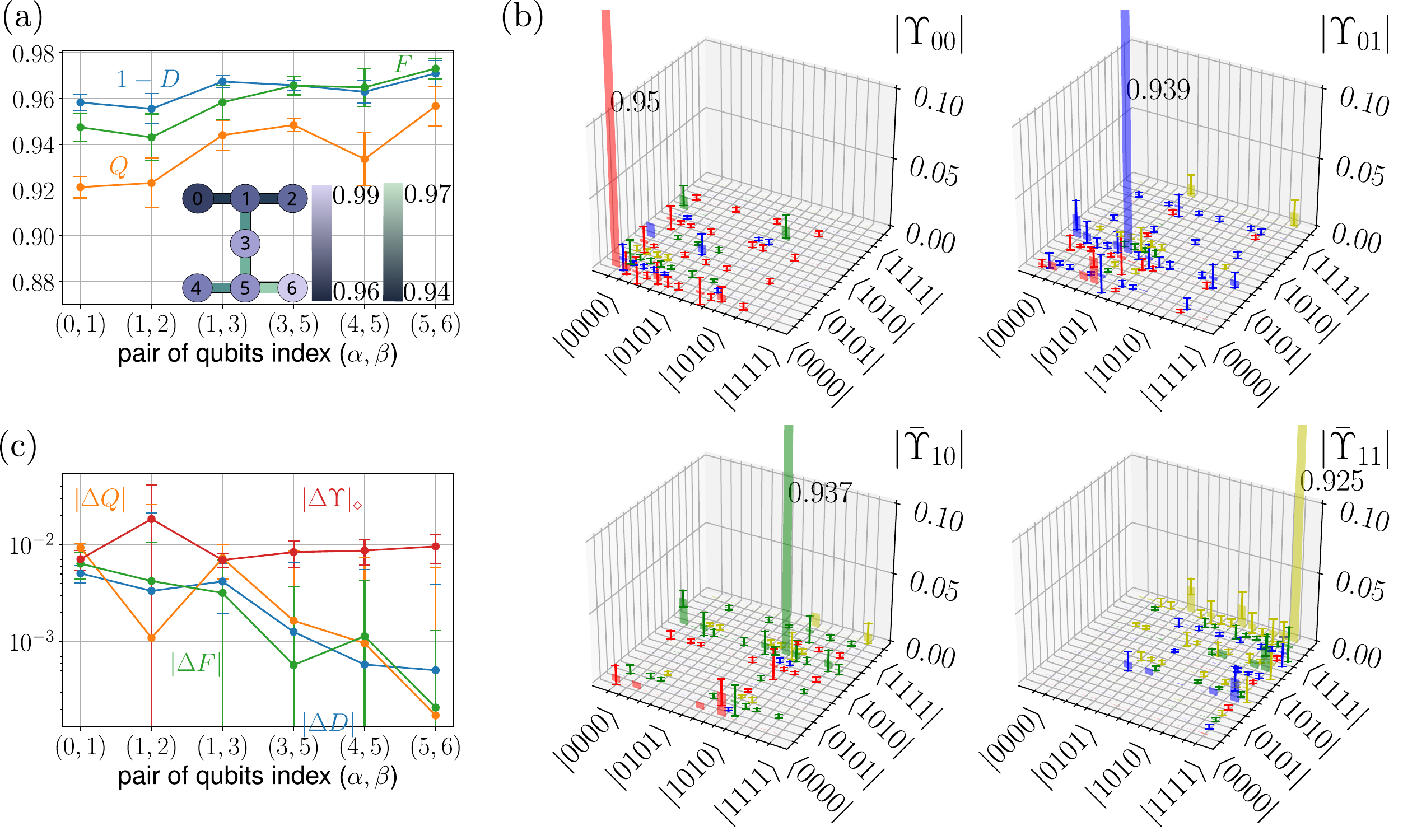}
	\caption{{\bf Experimental tomographic characterization of two-qubit measurements over every pair of connected qubits \lp{in parallel}.} (a) Measurement quantifiers $F$, $Q$, and $1-D$ for each nearest neighbor pair.\lp{ The inset represents} the average performance \lp{$(F+Q+1-D)/3$} in the device of each qubit (blue color code in circles) and of every pair of connected qubits (green color code in bars). (b) Reconstructed two-qubit Choi matrices $\bar{\Upsilon}_{nm}$ averaged over all connected pairs of qubits for the four possible outcomes with $n,m=0,1$. \lp{(c) Error in the quantifiers $F$, $Q$, and $1-D$ and the Choi operators introduced by the parallelization. Error bars are the standard deviation estimated with 5 realization of the experiment.} }
	\label{fig3}
\end{figure*}

\lp{The Choi matrices not only provide individual qubit metrics $(F, Q, D)$ but also hint the physical processes behind measurement errors. The Choi matrix element $p_{n}^{a\rightarrow b}=\braket{bb|\Upsilon_{n}|aa}$
quantifies the probability that the state flips from $\ket{a}$ to $\ket{b}$ when outcome $n$ is detected. Each element informs about deviations from the ideal projective measurement $p^{a\rightarrow a}_{a}=1$, as well as possible origins for those deviations.}

\lp{Let us first put this into practice} using the averaged Choi matrices $\bar{\Upsilon}_n=\sum_\alpha \Upsilon_n^\alpha/N$, shown in Figure~\ref{fig2}(b). Note how the readout of the $\ket{0}$ state $(p_0^{0\rightarrow0}=\lp{0.975})$ is implemented with a better quality than that of $\ket{1}$ $(p_1^{1\rightarrow1}=\lp{0.960})$. Bit flip noise is identified as the main source of errors, dominated by the qubit decay process $\ket{1}\rightarrow\ket{0}$ ($p_1^{1\rightarrow 0}=\lp{0.02}$), and slightly less influenced by the excitation channel $\ket{0}\rightarrow \ket{1}$ $(p_1^{0\rightarrow1}=\lp{0.016})$. Considering that \lp{\emph{\perth}} has a relaxation time $T_1\approx100\mu s$ and a measurement time $T=\lp{700}ns$~\cite{ibmq}, we estimate a \lp{baseline probability of qubit relaxation} $p_{\lpv{\rm th}}=1-e^{-T/T_1}\approx\tr{0.007}$, \tr{which accounts for 35$\%$ of the observed decay error. The remaining bit-flip error may be due to Purcell-induced decay and other non-dispersive errors that occur during the measurement process itself \cite{pereira_qnd_tomography}.}

This analysis may also be done qubit by qubit. Figures~\ref{fig2}(c)-(d) show the experimental Choi matrices for qubits $\alpha=6$ and $\alpha=0$, respectively (all others being included in \lpv{Supplementary Figures 1}). Qubit $\alpha=6$ is the best on the device, with \lp{$F=0.991$, $Q=0.985$, and $1-D=0.988$}. Its Choi matrices are also the closest ones to an ideal measurement, with large values of $p_{n}^{n\rightarrow n}$ and below average flip errors. On the other hand, Qubit $\alpha=0$ is the one exhibiting the worst performance with \lp{$F=0.971$, $Q=0.951$, and $1-D=0.97$}. The Choi matrices for this qubit are dominated by a strong bit flip error $p_1^{1\rightarrow0}=0.035$ and $p_1^{0\rightarrow1}=\tr{0.021}$ in the outcome $\ket{1}$, and non-dispersive errors given by \tr{elements $\braket{ab|\Upsilon_{0}|00}$ and $\braket{ab|\Upsilon_{1}|11 }$, with $a\neq b$}. The projection in this outcome is thus not done correctly, which explains the reduction in the QND-ness and indestructiveness.

The parallelized tomography of the qubits has obvious performance advantages, but it could increase the error of the operations~\cite{rudinger_experimental_2021}. To quantify potential deviations, we have compared the outcome of parallel tomography on \emph{\perth} with the independent characterization of those qubits, running the $\mathcal{O}(N)$ circuits separately. As shown in Fig~\ref{fig2}(e), the differences in the three quantifiers---fidelity $|\Delta F|=|F_{ind}-F_{par}|$, QND-ness $|\Delta Q|=|Q_{ind}-Q_{par}|$, and destructiveness $|\Delta D|=|D_{ind}-D_{par}|$---lay below $10^{-2}$, and are smaller than the non-idealities of those quantifiers (see Fig.~\ref{fig2}(a)). Similarly, we have quantified the average distance in diamond norm \cite{Benenti_2010,BlumeKohout2017} between the Choi operators computed using both strategies $|\Delta\Upsilon|_\diamond=|\Upsilon_{ind}-\Upsilon_{par}|_\diamond\leq 1$, and these lay below $1.4\times 10^{-2}$ (see  Fig.~\ref{fig2}(e)), validating the use of the parallelized strategy.

\subsection{\lp{Two-qubit QND measurement tomography and cross-talk quantification}}

The low distinguishability between parallel and independent single-qubit QND-MT suggests that measurement correlations are weak across the device. We can further quantify such correlations comparing the joint measurement process for pairs of qubits $(\alpha,\beta)$, given by  $\Upsilon^{\alpha\beta}_{mn}$ for outcomes $m,n=0,1$, with the individual measurement processes $\Upsilon^{\alpha}_{m}\otimes\Upsilon^{\beta}_{n}$.

The two-qubit QND-MT requires the evaluation of $324$ circuits---two measurement processes interspersed by layers of gates $V_i$ and $U_j$ (cf.~Fig.~\ref{fig1}(d)). Since characterizing all $N(N-1)/2$ pairs on an $N$-qubit device is very costly, we first focused on neighboring qubits, which we expect to exhibit the greatest correlations. More precisely, for a device with $M$ physical connections  $(\alpha,\beta)\in\mathcal{C}$ of $M$, we aim to reconstruct the $4M$ two-qubit Choi matrices.

This two-qubit QND-MT can be parallelized by executing similar circuits on non-overlapping pairs of physically connected qubits. This requires dividing the quantum processor into sets of edges that do not share a common qubit. For the 7-qubit \lp{\emph{\perth{}}} and the 65-qubit \emph{\brooklyn{}} quantum processors, illustrated in Fig.~\ref{fig1}(e)-(f), we only need three sets. For a generic planar graph with $M$ vertices, coloring theorems~\cite{Lewis2021} ensure that the number of sets is never larger than 4, \lp{setting a bound on the number of circuits $4\times 18^2$ that does not grow with the processor's size. Finally, the protocol requires solving $5M$ optimization problems,} a task that can be efficiently parallelized on classical computers. Here, we employ a parallel two-qubit QND-MT to characterize the readout of physically connected qubits on the IBM quantum device \tr{\emph{\perth}}. The experiment runs approximately in \tr{$38$} minutes --using \tr{$2^{13}$} shots per circuit-- and the post-processing in $3$ minutes.

Fig.~\ref{fig3}(a) shows the experimental results for the quantifiers $F$, $Q$, and $1-D$ describing the two-qubit measurement. \lp{We see an overall decrease in the readout performance of pairs qubits with respect to the single-qubit results in Fig.~\ref{fig2}(a), but still we identify the same qualitative behavior: $F$ and $Q$ increase from top to bottom of the device (see inset of Fig.~\ref{fig3}(a)) and QND-ness is the worst and most fluctuating quantifier. As discussed below, the two-qubit quantifiers of all connected pairs are very well approximated as products of the single-qubit ones---e.g. $F^{\alpha\beta}\approx F^{\alpha}F^{\beta}$ and $Q^{\alpha\beta}\approx Q^{\alpha}Q^{\beta}$---.} This explains the reduction in average two-qubit fidelity and QND-ness between pairs to $\bar{F}=\tr{0.958}$ and $\bar{Q}=\tr{0.937}$. Indestructiveness $1-D$ is the most stable quantifier, ranging between $\tr{0.955}$ and $\tr{0.97}$, but reduces its average in a similar amount to $1-\bar{D}=\tr{0.963}$.

\begin{figure}
	\centering
	\includegraphics[width=0.9\linewidth]{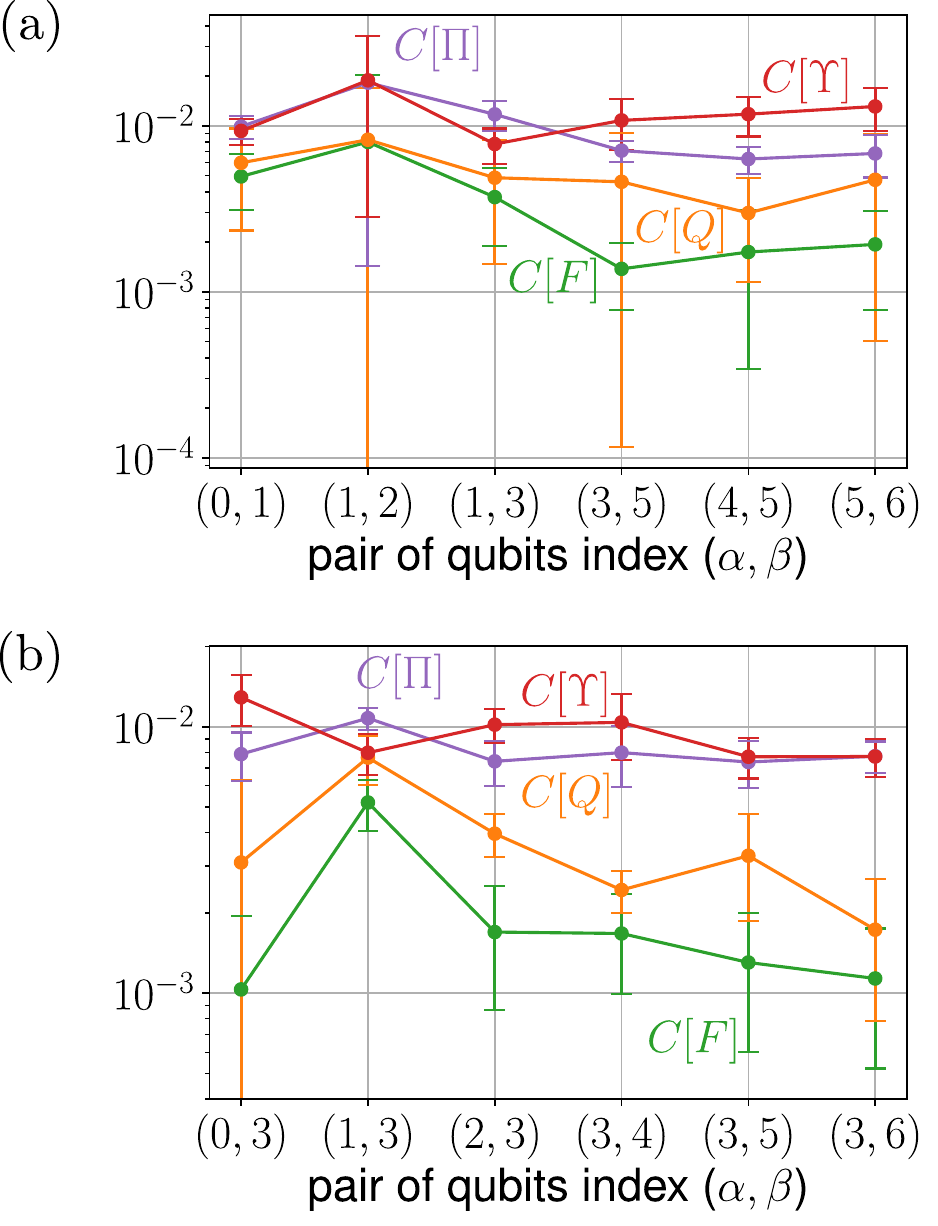}
	\caption{ \lp{{\bf Correlations in the joint measurement of pairs of qubits obtained by QND-MT.} (a) Correlations in fidelity $C[F]$, QND-ness $C[Q]$, and Choi operators $C[\Upsilon]$ for the joint readout of each pair of physically connected qubits. (b) Same Correlations $C[F]$, $C[Q]$, and $C[\Upsilon]$ for the readout of qubit $\alpha=3$ with every other qubit of the device.}}
	\label{fig4}
\end{figure}

Figure~\ref{fig3}(b) shows the two-qubit Choi matrices averaged over all pairs of connected qubits, $\bar\Upsilon_{nm}=\sum_{(\alpha,\beta)\in\mathcal{C}}\Upsilon_{nm}^{\alpha\beta}/M$. \lpv{The Choi matrices for all pairs are included in Supplementary Figures 2}. The largest probabilities of type $p_{ab}^{ab\rightarrow ab}=\braket{ab|\bar{\Upsilon}_{ab}|ab}$ show the same behavior as $p_{a}^{a\rightarrow a}=\braket{a|\Upsilon_a|a}$ for the singe-qubit case. The lowest deviation from the ideal measurement corresponds to the state $\ket{00}$ ($p_{00}^{00\rightarrow00}=\tr{0.95\approx (p_{0}^{0\rightarrow0})^2}$), followed by $\ket{01}$ and $\ket{10}$ ($p_{01}^{01\rightarrow01}= p_{10}^{10\rightarrow10}=\tr{0.938\approx p_{0}^{0\rightarrow 0}p_{1}^{1\rightarrow 1}} $), and finally the two-excitation state $\ket{11}$ (with $p_{11}^{11\rightarrow11}=\tr{0.924\approx (p_{1}^{1\rightarrow 1})^2}$), \tr{which suffers more from bit-flip errors.}

\lp{As in the single-qubit case, we estimate the errors introduced by parallelization by comparing the parallelized two-qubit QND-MT with the independent tomography of each pair. Fig.~\ref{fig3}(c) shows the error in fidelity, QND-ness, destructiveness, and Choi operators for each pair of physically connected qubits, as defined in the previous section. Parallelization introduces an error below $2\times 10^{-2}$ in all quantifiers and Choi operators.}

We quantify the measurement crosstalk by comparing the measurement processes of individual and pairs of qubits. This is done at the level of quantifiers, introducing heuristic measures of separability for the fidelity $C[F^{\alpha\beta}]=|F^{\alpha\beta}-F^{\alpha}F^{\beta}|$ and for the QNDness $C[Q^{\alpha\beta}]=|Q^{\alpha\beta}-Q^{\alpha}Q^{\beta}|$. It is also done at the level of operators, with estimates of the POVM correlation $C[\Pi^{\alpha\beta}]$ and the Choi correlation $C[\Upsilon^{\alpha\beta}]$ (see methods).

\lp{As hinted above, we observe a good separability of quantifiers. In Fig.~\ref{fig4}(a) we see correlations of \emph{\perth} device below $10^{-2}$ for all pairs, allowing us to estimate the fidelity and QNDness of pairs of qubits as products of the properties of individual qubits.}

\lp{Figure~\ref{fig4}(a) shows the POVM and Choi correlations for the \emph{\perth} device. We} certify the presence of measurement cross-talk between all physically connected pairs of qubits: \lp{all POVM elements} and Choi matrices are non-separable with correlations on the order of \lp{$10^{-2}$}, which exceed the statistical error bars from the tomography for \lp{most of the qubits. This represents a crosstalk error of about $1\%$, which is smaller than the physical error found on single- and two-qubit tomography.}

\lp{QND-MT is not restricted to nearest-neighbor correlations. As example, we have analyzed the correlations between all qubits in \emph{\perth{}} and the central qubit $\alpha=3$, in 6 sets of separate experiments. This produces pairs at two different distances, the first neighbors $(1,3)$ and $(3,5)$, and the second neighbors $(0,3)$, $(2,3)$, $(3,4)$, and $(3, 6)$. Fig.~\ref{fig4}(b) shows the correlations obtained for those pairs. We can see that correlations $C[F^{\alpha\beta}]$ and  $C[Q^{\alpha\beta}]$ are of order $10^{-3}$ for all qubits, while $C[\Pi^{\alpha\beta}]$ and $C[\Upsilon^{\alpha\beta}]$ are approximately $10^{-2}$. In this small device we do not observe a clear decay of correlations with distance, but we verify that all correlations are smaller than the measurement errors detected for independent single-qubit tomography.}

\subsection{\lp{Scaling of QND-MT on larger devices}}

\lp{Parallel single-qubit QND-MT is an efficient technique to characterize large devices, that requires a fixed number of circuits---$82$ including GST---independently of the device size. Using the execution times obtained in the experiments on the \emph{\perth} we can extrapolate the performance in larger devices.} For the 65-qubit \emph{\brooklyn}, with a degree 3 connectivity shown in Fig. \ref{fig1}f and a smaller CLOPS number of $1.5 \times 10^3$ \cite{clops}, we \tr{estimate} $4$ minutes for the single-qubit characterization and \tr{$5$ minutes of post-processing in a Ryzen-7 5800H processor with 8 cores}. Notice that all experimental execution times do not depend on the size of the device but they are limited by the number of CLOPS, which are typically lower for larger devices.

\lp{We have discussed also three strategies to certify the errors in parallel QND-MT. One strategy is the application of QND-MT of individual qubits in separate, non-parallel experiments. This has a cost that grows linearly $\mathcal{O}(N)$ with the number of sampled qubits, but it is a routine that may be applied with less frequency than the complete calibration. This method enables the development of heat maps of the chip and suggests the order of magnitude of underlying correlations.}

\lp{The second strategy is the parallelized QND-MT of pairs of neighboring qubits, a method that will provide results consistent with the previous methodology, but also give information about the strength of the cross-talk. In the two-qubit parallelized strategy, our estimate give a total of $1296$ independent circuits for any device size, taking $63$ minutes for the two-qubit circuit evaluation in a 65-qubit \emph{\brooklyn} processor, and $30$ minutes in a Ryzen-7 5800H processor with 8 cores.}

\lp{The third and most expensive strategy is to implement a two-qubit QND-MT for \textit{all qubit pairs} in large devices with $O(\log(N))$ parallel groups~\cite{Cotler2020overlapping} and $O(N^2)$ optimization problems. In this case, we estimate $2.5$ hours for the experiment and a similar amount of post-processing to characterize the \emph{\brooklyn} device. This is an efficient scaling that enables a very robust calibration of the complete device, to be done only sporadically.}

\lp{Finally, for larger devices, the execution and post-processing times could be too long for a complete two-qubit measurement tomography---extending to days for devices with more than $1000$ qubits---in which case it makes sense to either randomly sample those pairs, or concentrate the study to specific regions of the chip, that revealed more problematic in the first two methods.}

\subsection{\lp{QND measurement tomography of generalized measurements}}

\begin{figure*}[t]
	\centering
	\includegraphics[width=0.8\linewidth]{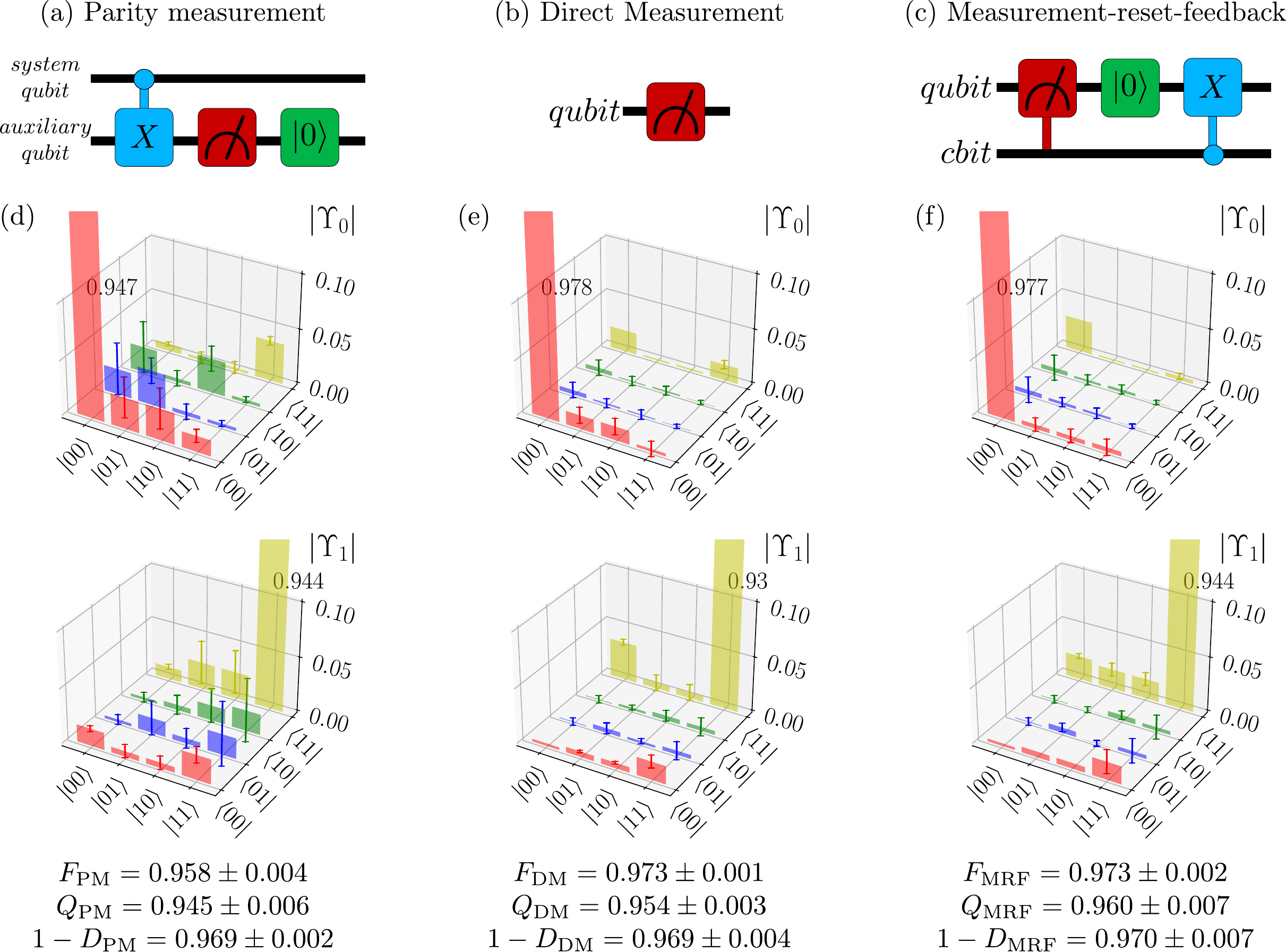}
	\caption{{\bf Experimental QND-MT characterization of \lp{generalized measurements}.} Circuits depicting (a) \lp{a single-qubit parity measurement via an auxiliary qubit}, (b) a direct qubit measurement, and (c) a measurement-and-reset scheme \lp{with classical feedback}. (d)-(f) Choi matrices and quantifiers \lp{corresponding to each scheme. Error bars are the standard deviation estimated from 5 realization of the experiments}.}
	\label{fig5}
\end{figure*}

\lp{The QND-MT protocol we introduce can be applied to} any kind of generalized measurements~\cite{breuer_theory_nodate}. These include synthetic measurements that combine standard detectors with other computing elements, such as local and entangling gates, \lp{auxiliary} qubits, and resets.

\lp{In this work we discuss the application to stabilizer measurements, a relevant example which are widely used in quantum error correction protocols~\cite{krinner_realizing_2021,google_error_correction_2021,Zhao2021surfacecode}. Such measurements are usually implemented with controlled operations over an auxiliary qubit, which is finally measured and reset, to discriminate states with different stabilizer value. If we trace over the auxiliary qubit, the generalized measurement is, up to implementation errors, QND, enabling the repetitive monitoring of error syndromes.}

As an illustration of how QND-MT works with a generalized measurement, we discuss a single-qubit parity measurement (PM). As shown in Fig.~\ref{fig5}(a), this protocol includes an auxiliary qubit, a controlled CNOT operation, and a single-qubit readout and reset. \lpv{Note that, unlike all higher parity measurements, the single-qubit PM
does not entangle multiple system qubits and thus it is not directly applicable to quantum error correction codes. However, it already includes all the underlying operations supporting multi-qubit PM, which can be scaled to characterize multi-qubit measurement errors in practical error correction codes. Here, we study the performance of the single-qubit PM} using two fixed qubits of the \emph{\perth} quantum processor, \lpv{and we compare it} with the performance of the direct measurement (DM) on the same system qubit, as shown in Fig.~\ref{fig5}(b). The Choi matrices and the quantifiers obtained for the parity and direct measurements are shown in Figs.~\ref{fig5}(d) and (e), respectively.

\lp{In this study we observe a decrease of the fidelity and QND-ness of the PM with respect to the DM. The readout fidelity of the parity measurement $F_{\lpv{\rm PM}}=0.958$ is close to the product of $F_{\lpv{\rm DM}}=0.973$ and the fidelity of the CNOT provided by IBM $F_{\lpv{\rm CNOT}}=0.9897$. Therefore, we can conclude that this decrease is mainly due to the CNOT gate as the error from reset is expected to be smaller than $1\%$ \cite{PhysRevLett.121.060502}. The indestructiveness is the same for parity and direct measurements, $1-D_{\lpv{\rm PM}}=1-D_{\lpv{\rm DM}}=0.969$, which is consistent with the fact that the CNOT and reset operations do not add measurement back-action on the system. In the Choi operators, we can also see the appearance of new bars that describe the noise introduced by the CNOT gate, as well as an increase in the overall error bars and fluctuations. }

\lp{Another interesting example of generalized measurement is} the measure-reset-\lp{feedback} (MRF) operation, shown in Fig.~\ref{fig5}(c). \lp{It consists of} a QND measurement followed by a reset and a classically-conditioned NOT operation that brings the measured qubit exactly to the quantum state \lp{selected as measurement outcome---i.e. the qubit is reset to state $\ket{0}$ or $\ket{1}$ when the measurement outcome was deemed $n=0$ or $1$, respectively}. If the reset and NOT operations have high fidelities, measurement-and-reset should \lpv{fix} the QND nature of a measurement, bringing the errors $1-Q$ and $D$ closer to the measurement infidelity $1-F$.

We \lp{applied QND-MT to this generalized measurement using a single qubit of the IBM-Q \emph{\perth} processor. The resulting Choi matrices and quantifiers are shown in Figure~\ref{fig5}(f). The MRF scheme has better performance than the DM in the same qubit, having approximately the same fidelity $F_{\lpv{\rm DM}}\approx F_{\lpv{\rm MRF}}\approx 0.975$ and indestructiveness $1-D_{\lpv{\rm DM}}\approx 1-D_{\lpv{\rm MRF}}\approx 0.969$, but with an increase of QNDness from $Q_{\lpv{\rm DM}}=0.954$ to $Q_{\lpv{\rm DM}}=0.960$. Considering the error bars of the QND-ness and indestructiveness, we find that the worst case MRF provides a QND readout with the same quality as a direct measurement. Moreover, we also witness a reduction in the noisy components of the Choi operator, such as those describing bit-flip errors. }

\section{Discussion}

In this work we have demonstrated an efficient, highly parallelizable protocol for QND measurement tomography of a state-of-the-art \tr{multi-qubit quantum computer, which works with both single-qubit measurement, as well as generalized measurements---e.g. error syndrome measurements, parity measurements, etc.} Our method is based on a self-consistent reconstruction of the Choi matrices for single-qubit and two-qubit measurements, which provides information about measurement quality, the QND nature of the measurement and the strength and type of errors.

\lp{In the single-qubit scenario, we have developed strategies to massively parallelize the tomography, an approximation that works when multiple measurements can be executed with small crosstalk or correlation.} We have applied this protocol in experiments with a 7-qubit IBM quantum computer, obtaining fascinating insight into the performance of the device. First of all, we have found that the chip is well tuned to high-fidelity measurements, with weak and long pulses---much longer than single- or two-qubit gates---that mitigate non-dispersive and discrimination errors, at the expense of increasing incoherent errors, in particular single-qubit bit flip. This limits the QND nature of the measurement which fluctuates along the different qubits of the device.

\lp{We have also developed different strategies determine whether single-qubit measurements are independent, and can be parallelized. The most sophisticated strategy involves applying QND-MT to the simultaneous measurement of two qubits, to reconstruct the joint Choi matrices and quantify the degree of correlation. In the setup considered, these correlations lay below 1$\%$ and validate the parallelization strategy which, as discussed above, can be efficiently scaled to large multi-qubit processors with an almost fixed cost.}

Finally, we have also demonstrated how QND-MT can be generalized to custom measurements, in particular \lp{to parity-type measurements relevant to quantum error correcting codes and measurement-and-reset schemes with classical feedback. We used the Choi matrices to identify coherent errors introduced by the CNOT gate in parity measurements and we provided evidence that the reset operation with classical feedback is an appealing way to improve the QND quality of a measurement.}

This work opens several avenues for further research. The obvious one is to use QND-MT as an input for systematic optimization of the measurement pulses. The goal here is to optimize the driving amplitude and measurement time, minimizing the errors that manifest in the Choi matrices. This would allow us to reduce the decay channels found in the experiment, while keeping other sources of error at bay---e.g. non-dispersive effects~\cite{govia_entanglement_2016}, discrimination errors~\cite{Martinez2020improving,pereira_qnd_tomography}, decoherence~\cite{boissonneault_dispersive_2009}, leakage to higher levels of the transmon~\cite{wang_optimal_2021}, or rotating wave corrections~\cite{Sank2016}. Another approach is to design alternative schemes for qubit readout that may be more QND~\cite{Didier2015fast,Dassonneville2020fast,Touzard2019gated}, but this would add new error sources that could be similarly identified and characterized with the application of QND-MT. 

\lp{An additional} research avenue is to further understand and mitigate the correlations between simultaneous measurements. In this work we have explored two-qubit correlations, but higher-order correlations, involving 3 or more qubits, could also be analyzed with the help of better tomography methods, such as compressed sensing~\cite{gross2010cs,Riofro2017experimental}. \tr{These methods could also be used to quantify readout and cross-talk errors occurring in multi-qubit stabilizer measurements involving plaquettes of 4 or more qubits as they are required in practical quantum error correction codes such as the surface~\cite{krinner_realizing_2021,google_error_correction_2021,Zhao2021surfacecode} or color codes \cite{postler_demonstration_2022}.} 

\section{Methods}

\subsection{\lp{QND measurement quantifiers}}

\lp{To characterize the most important properties of non-destructive measurements we employ three quantifiers: the readout fidelity, the QNDness, and the destructiveness. Here, we show how to obtain these quantifiers from the reconstructed Choi matrices as introduced in~\cite{pereira_qnd_tomography}.}

The fidelity $F$ is the standard quantifier of a detector's readout performance, measured by the probability that an initially prepared eigenstate $\ket{n}$ is successfully identified,
\begin{equation}
  F = \frac{1}{2^N} \sum_{n=0}^{2^N-1} \braket{ n|\Pi_n|n }.
\end{equation}
The fidelity can be interpreted as the efficiency of the readout as it can be related to the signal-to-noise ratio of the measurement~\cite{blais_circuit_2020,Didier2015fast}. It ignores any information about the post-measurement state and the QND nature of the measurement. \lp{The QND-ness $Q$ incorporates information from the post-measurement state and quantifies how close are the Choi matrices with respect to an} ideal projective measurement. In quantitative terms, it is the probability that an initially prepared eigenstate $\ket{n}$ is preserved and successfully identified in two consecutive measurements,
\begin{equation}
  Q = \frac{1}{2^N} \sum_{n=0}^{2^N-1} \braket{ nn|\Upsilon_n|nn }.
\end{equation}
The destructiveness $D$~\cite{pereira_qnd_tomography} asserts precisely the QND nature of generic measurements by verifying the preservation of the expectation value $\langle O\rangle$ after the measurement. Operationally, it is defined as the largest change suffered by any observable compatible with $O$ as,
\begin{equation}
    D = {} \frac{1}{2}\max_{||O_c||=1}\Vert O_c - \mathcal{E}^\dagger(O_c)\Vert,\quad [O,O_c]=0, \label{EqD}
\end{equation}
where $\Vert\cdot\Vert$ is the \lp{Hilbert-Schmidt} norm. Unlike $F$ and $Q$, computing $D$ requires a complete tomographic reconstruction of the measurement process, $\mathcal{E}^\dagger(O_c)= \sum_{ijkln}\left[\Upsilon_n^{klij}\right]^\ast O_c^{kl}|i\rangle\langle j|$, but it allows us to quantify the measurement back-action without the bias of $Q$ towards ideal measurements~\cite{pereira_qnd_tomography}. \lpv{Note that equation \eqref{EqD} is motivated by the definition of a QND measurement, $\Tr(O\rho)=\Tr(O\mathcal{E}(\rho))$. Moving into the Heisenberg picture this condition becomes $\Tr(O\rho)=\Tr(\mathcal{E}^\dagger(O)\rho)$, where $\mathcal{E}^\dagger$ is the self-adjoint process of $\mathcal{E}$. Therefore, we can quantify how QND is a measurement by the deviation between $O$ and $\mathcal{E}^\dagger(O)$, that is $||O-\mathcal{E}^\dagger(O)||$. The last step to obtain Eq.~\eqref{EqD} consists in searching for the largest disagreement over the set of all normalized observables compatible with $O$, so that we ensure that $D$ is an upper bound for the back-action of the measurement.}

\subsection{\lp{Quantification of measurement cross-talk and correlations}}

\lp{To quantify the correlations in the measurement of pairs of qubits we introduce heuristic metrics that compare the POVM and Choi matrices derived from two-qubit tomography with the tensor product of the operators obtained from single-qubit tomography. Although this is a comparison between two detector models rather than a intrinsic property of the operators, the outcome provides information about the operation of the device and the effect of including higher order interactions in the QND-MT. We also use these correlations to quantify the distinguishability error of performing the tomography in parallel or independently, as shown below.}

\lp{First, we define the correlation in two-qubit Choi operators $C[\Upsilon^{\alpha\beta}]$.} Let $\Upsilon_n^\alpha$ and $\Upsilon_m^\beta$ be the process operators of two single qubits $\alpha$ and $\beta$, and let $\Upsilon_{nm}^{\alpha\beta}$ be the joint measurement of both qubits. We define the Choi matrices correlation as
\lp{\begin{align}
   C[\Upsilon^{\alpha\beta}] = \frac{1}{8} \sum_{nm} || \Upsilon_{nm}^{\alpha\beta} - \Upsilon_n^{\alpha}\bar\otimes\Upsilon_m^{\beta}||_{\diamond}\label{CUpsi}
\end{align}}
where $\bar{\otimes}$ is tensor product operation in the super-operator space and \lp{$||\cdot||_\diamond$ is the diamond norm \cite{Benenti_2010,BlumeKohout2017}. This quantity not only evaluates the distance between the processes $\Upsilon_n^\alpha\bar{\otimes}\Upsilon_m^\beta$ and $\Upsilon_{nm}^{\alpha\beta}$, but is also related to the probability of discriminating the quantum states generated by them. This probability is given by $P_{d}=(1 + C[\Upsilon^{\alpha\beta}] )/2$. Therefore, a small $C[\Upsilon^{\alpha\beta}]\ll 1$ means that the post-measurement states are nearly indistinguishable. Notice that the pre-factor in $C[\Upsilon^{\alpha\beta}]$ are chosen normalize the correlation between 0 and 1.}

\lp{Conveniently, we can use the same definition in Eq.~(\ref{CUpsi}) to evaluate the distinguishability of Choi matrices reconstructed in parallel $\Upsilon_{par}$ or independently $\Upsilon_{ind}$ as $C[\Delta \Upsilon]=C[\Upsilon_{par}-\Upsilon_{ind}]$, and thereby quantify the error introduced by the parallelization. This is done in Figs.~\ref{fig2}(e) and \ref{fig3}(c).}

\lp{In a similar spirit as done for $C[\Upsilon^{\alpha\beta}]$, we can define the correlation $C[\Pi^{\alpha\beta}]$ in the POVMs of two-qubit measurements.} Let $\Pi_n^\alpha$ and $\Pi_m^\beta$ be the POVM elements of two single qubits $\alpha$ and $\beta$, and $\Pi_{nm}^{\alpha\beta}$ be the joint POVM element of both qubits. We define the POVM correlation as
\lp{\begin{align}
    C[\Pi^{\alpha\beta}] = \frac{1}{4}\sum_{nm}||\Pi_{nm}^{\alpha\beta} -\Pi_n^\alpha\otimes\Pi_m^\beta||_2,
\end{align}}
\lp{where $||\cdot||_2$ is the 2-norm, that is, the largest singular value. This quantity establishes an upper bound for the average error on the probability distribution predicted by the single-qubit reconstruction $P_{nm}^S = Tr( \rho [\Pi_n^\alpha\otimes\Pi_m^\beta])$ compared with the joint measurement $P_{nm}^J = Tr( \rho \Pi_{nm}^{\alpha\beta})$, given by $\sum_{nm}|P_{nm}^S-P_{nm}^J|/4\leq C[\Pi^{\alpha\beta}]$ for any density matrix $\rho$. Notice that $C[\Pi^{\alpha\beta}]$ is normalized between 0 and 1 as $C[\Upsilon^{\alpha\beta}]$.}

\subsection{Maximum Likelihood \tr{estimation for} QND measurement tomography}

Maximum likelihood estimation (MLE) is a statistical inference method widely used in quantum tomography. MLE allows us to recover density matrices, POVMs, or Choi matrices that are meaningful and satisfy all the physical constraints of a measurement. It achieves this goal by optimizing the likelihood function \lp{$\mathcal{L}(\theta|\hat{f})$ of the experimental data $\hat{f}$ for a given parametric model $\mathcal{M}(\theta)$. We employ as a Gaussian distribution as a likelihood function, 
\begin{equation}
    \mathcal{L}(\theta|\hat{f}) = \sum_{i}\frac{ [\hat{f}(i) - p(i)]^2 }{ p(i) },\label{MLE}
\end{equation}
where $\hat{f}(i)$ are the estimated probabilities obtained from the experiment and $p(i)$ are the theoretical probabilities predicted by the model $\mathcal{M}(\theta)$. We minimize this likelihood function (\ref{MLE}) for both the QND-MT and GST. Notice that, for simplicity, the notation of the theoretical probabilities $p(i)$ omits the dependence on the parameters $\theta$, and that the index $i$ may refer to a group of indices as shown below.}

The QND-MT consists of two steps, first a measurement tomography of the POVM and then a process tomography of each Choi matrix. We reconstruct the POVMs $\{\Pi_j\}$ by first obtaining the \lp{theoretical probabilities,
\begin{equation}
    p(n|k) = \Tr(\Pi_n V_k(\rho)),
\end{equation}
of obtaining the outcome $n$ condition to the application of gate $V_k$. We then minimize the likelihood function of form (\ref{MLE}) over the} set of feasible matrices $\lbrace \Pi_n\rbrace$ satisfying $\Pi_n\geq 0$ and $\sum_n \Pi_n =\openone$. Finally, we estimate the Choi matrices $\Upsilon_n$ by obtaining the \lp{theoretical probabilities,
\begin{equation}
     p(mn|jk)= \Tr[(U_j^\dagger(\Pi_m) \otimes V_k(\rho)^T)\tilde\Upsilon_n],
\end{equation}
of obtaining the outcome $n$ in the second measurement and the outcome $m$ in the first measurement, condition to the application of gates $j$ and $k$. We then 
minimize the corresponding likelihood function of the form (\ref{MLE}) over the} set of the Choi matrix $\Upsilon_n$ that satisfies $\tilde\Upsilon_n\geq 0$ and the POVM constraint \lp{$\Tr_1(\tilde\Upsilon_n)=\Pi_n$. Here, $\Tr_1(\cdot)$ is the partial trace over the first subsystem, and} $\tilde\Upsilon$ is the transposed Choi matrix, \lp{which is a positive matrix with} elements $\braket{ik|\tilde\Upsilon|jl}=\braket{ij|\Upsilon|kl}$. 

\lp{To separate experimental errors in gates and state preparation from the measurement errors that we want to characterize, we can apply a GST previously to the QND-MT. The GST gives us an experimental estimate of the set $\{\rho, \Pi_i, G_j\}$, composed of estimators of the initial state, the POVM elements, and the gates, respectively. Here $\{\mathcal{G}_j\}$ are generic trace-preserving processes and not necessarily unitary operations. The theoretical probabilities of obtaining the outcome $l$ are 
\begin{equation}
    p(l|ijk) = \Tr(\Pi_l G_k G_j G_i(\rho) ), \label{probGST}
\end{equation}
which are condition to the application of gates $i,j,k$} as shown by the circuits in Fig.~\ref{fig1}(b). \tr{We then minimize (\ref{MLE}) by comparing the probabilities (\ref{probGST}) with the experimental data to obtain a physically meaningful set $\{\rho, \Pi_i, G_j\}$ that self-consistently accounts for state preparation, gates, and measurement errors. Notice that when we use GST, we can omit the first step of QND-MT as we already have an experimental estimate of the POVMs $\{\Pi_i\}$. In addition, the gates $\{U_j\}$ and $\{V_k\}$ needed for the second step of QND-MT must be formed as concatenations of the $\{G_k\}$ processes in order to account for gate errors.}

In total, QND measurement tomography of the device requires solving $3N+5M$ optimization problems. We solve them using sequential least-squares programming, satisfying the positivity of operators via Cholesky decomposition, and the completeness constraints via Lagrange multipliers.

\lp{To quantify the goodness-of-fit of our estimators, we employ the $\chi^2$-test~\cite{Langford2013,Temme2015}. This is a standard tool for statistical hypothesis testing, that is, for rigorously deciding if there is enough evidence to reject a model. In our work, we apply this test to all single- and two-qubit Choi matrix reconstructions and demonstrate that the fits and models are in agreement with the experimental data within a standard confidence interval of $95\%$. See \lpv{Supplementary Methods 1} for a detailed analysis and a description of the method.}   

\section{acknowledgments}

This work has been supported by funding from Spanish project PGC2018-094792-B-I00 (MCIU/AEI/FEDER, UE), CAM/FEDER Project No. S2018/TCS-4342 (QUITEMAD-CM), \tr{and the Proyecto Sinérgico CAM 2020 Y2020/TCS-6545 (NanoQuCo-CM)}. L.P. was supported by ANID-PFCHA/DOCTORADO-BECAS-CHILE/2019-772200275. T.R. further acknowledges support from the Juan de la Cierva fellowship IJC2019-040260-I. We thank the IBM Quantum Team for making multiple devices available to the CSIC-IBM Quantum Hub via the IBM Quantum Experience. The views expressed are those of the authors, and do not reflect the official policy or position of IBM or the IBM Quantum team.


%

\onecolumngrid
\newpage
\begin{center}
	\textbf{\large Supplementary Material:\\ 
		Parallel QND measurement tomography of multi-qubit quantum devices}
	\vspace*{0.25cm}
	
	L. Pereira$^{1}$, J.J.~Garc\'ia-Ripoll$^{1}$, and T. Ramos$^{1}$\\
	\vspace*{0.15cm}
	\small{\textit{$^1$ Instituto de F\'{i}sica Fundamental IFF-CSIC, Calle Serrano 113b, Madrid 28006, Spain}}
	\vspace*{0.25cm}
\end{center}

\twocolumngrid

\section{Supplementary figures}
\subsection{I.- All experimental single-qubit Choi matrices}

In this section, we show the absolute value of all the reconstructed single-qubit Choi matrices $\Upsilon_n^\alpha$ for all possible measurement outcomes $n=0,1$ of every qubit $\alpha=0,\dots,6$ of the \emph{ibm$\_$perth} chip shown in Figure~1(e) of the main text. All the definitions, calculation procedures, and error treatments are explained in the main text and methods.

\begin{figure}[H]
	\centering
	\includegraphics[width=\linewidth]{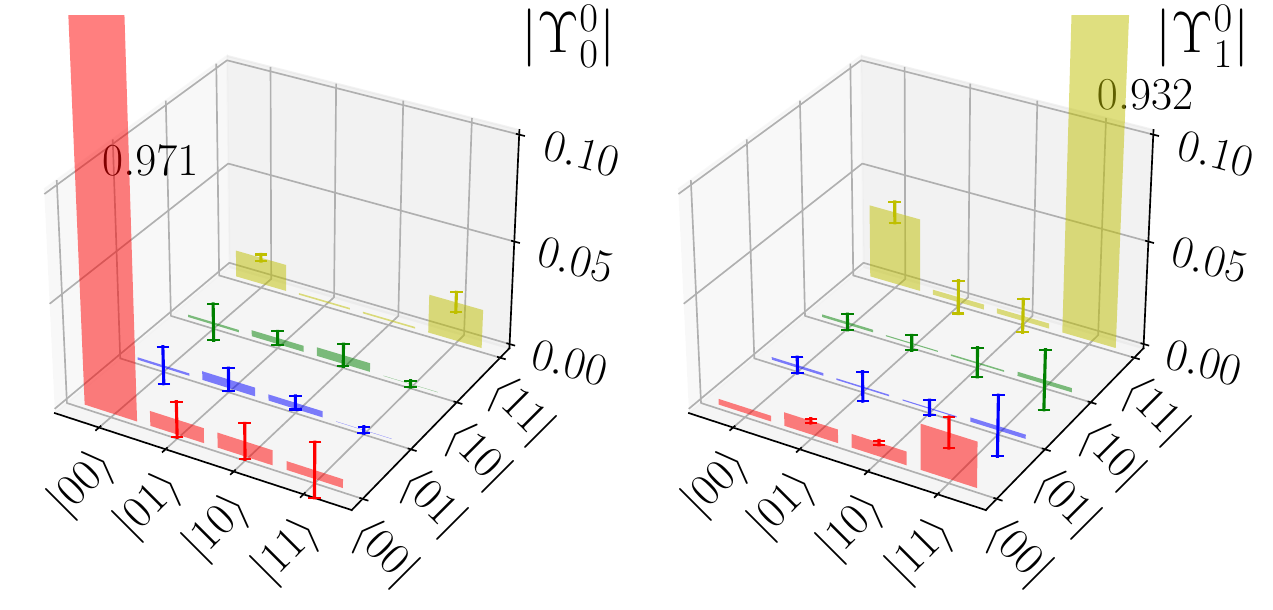}
	\caption{Absolute value of the reconstructed Choi matrices $\Upsilon_n^0$ for a measurement on qubit $\alpha=0$. Error bars are the standard deviation estimated with 5 realizations of the experiment.}
	\label{fig:q0}
\end{figure}

\begin{figure}[H]
	\centering
	\includegraphics[width=\linewidth]{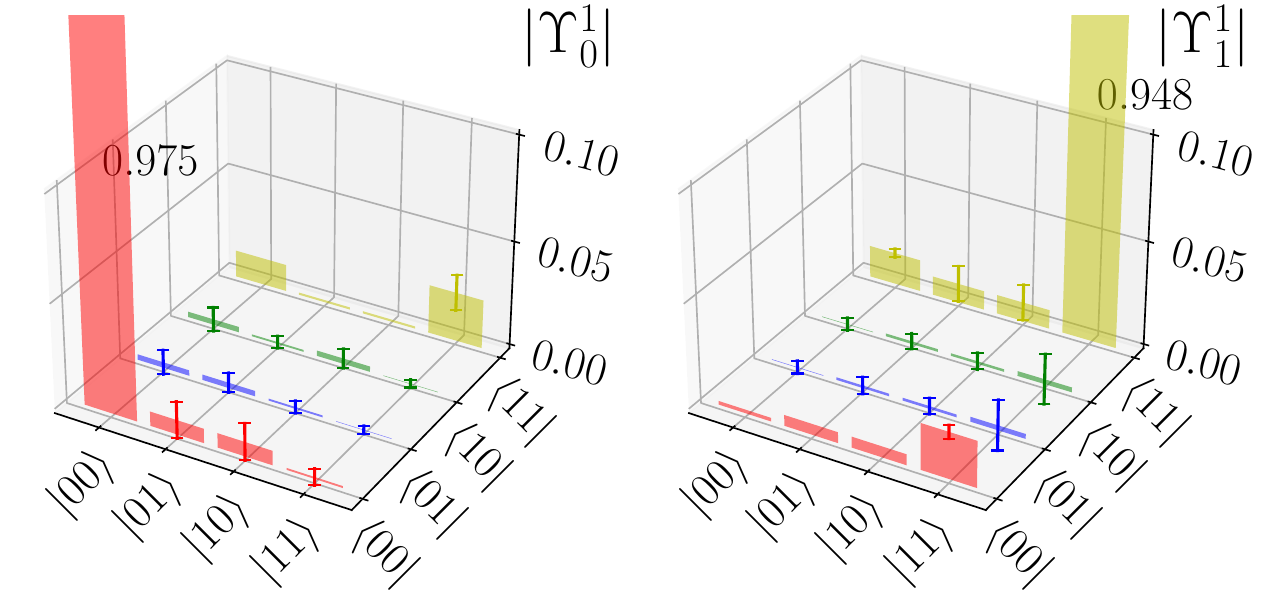}
	\caption{Absolute value of the reconstructed Choi matrices $\Upsilon_n^1$ for a measurement on qubit $\alpha=1$. Error bars are the standard deviation estimated with 5 realizations of the experiment.}
	\label{fig:q1}
\end{figure}

\begin{figure}[H]
	\centering
	\includegraphics[width=\linewidth]{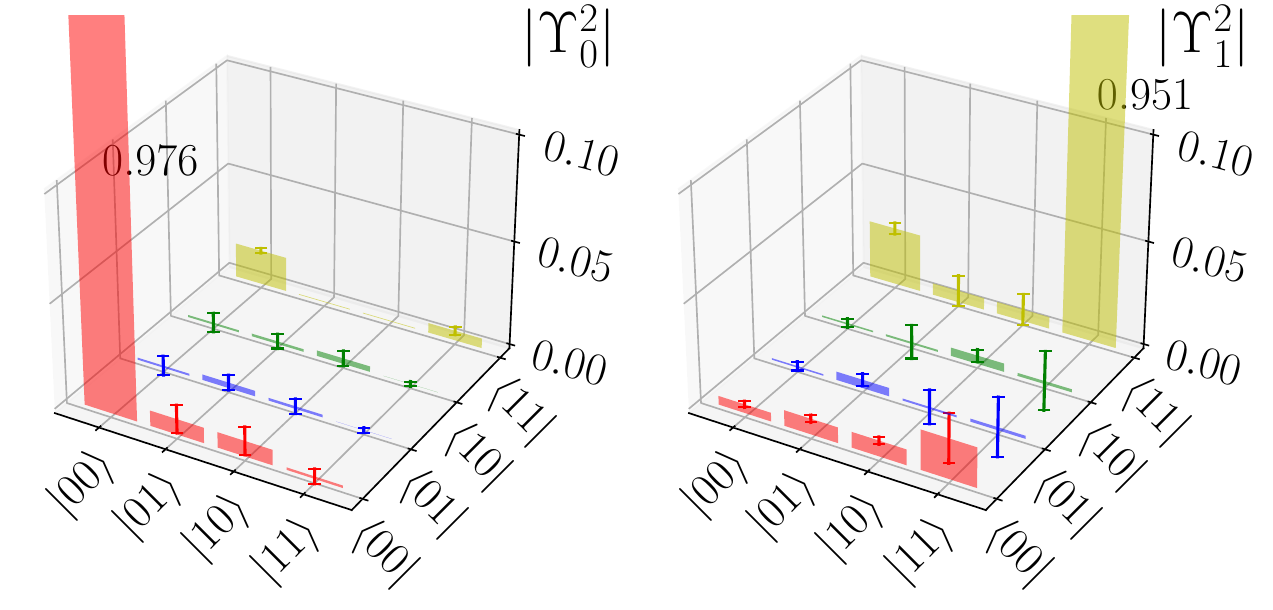}
	\caption{Absolute value of the reconstructed Choi matrices $\Upsilon_n^2$ for a measurement on qubit $\alpha=2$. Error bars are the standard deviation estimated with 5 realizations of the experiment.}
	\label{fig:q2}
\end{figure}

\begin{figure}[H]
	\centering
	\includegraphics[width=\linewidth]{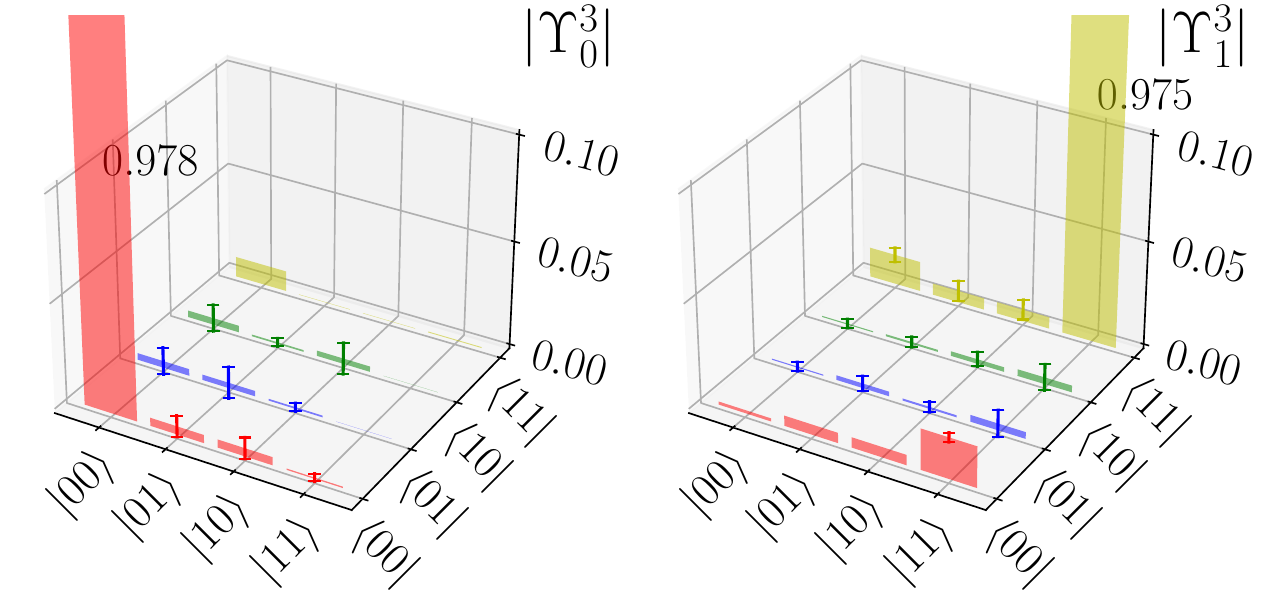}
	\caption{Absolute value of the reconstructed Choi matrices $\Upsilon_n^3$ for a measurement on qubit $\alpha=3$. Error bars are the standard deviation estimated with 5 realizations of the experiment.}
	\label{fig:q3}
\end{figure}

\begin{figure}[H]
	\centering
	\includegraphics[width=\linewidth]{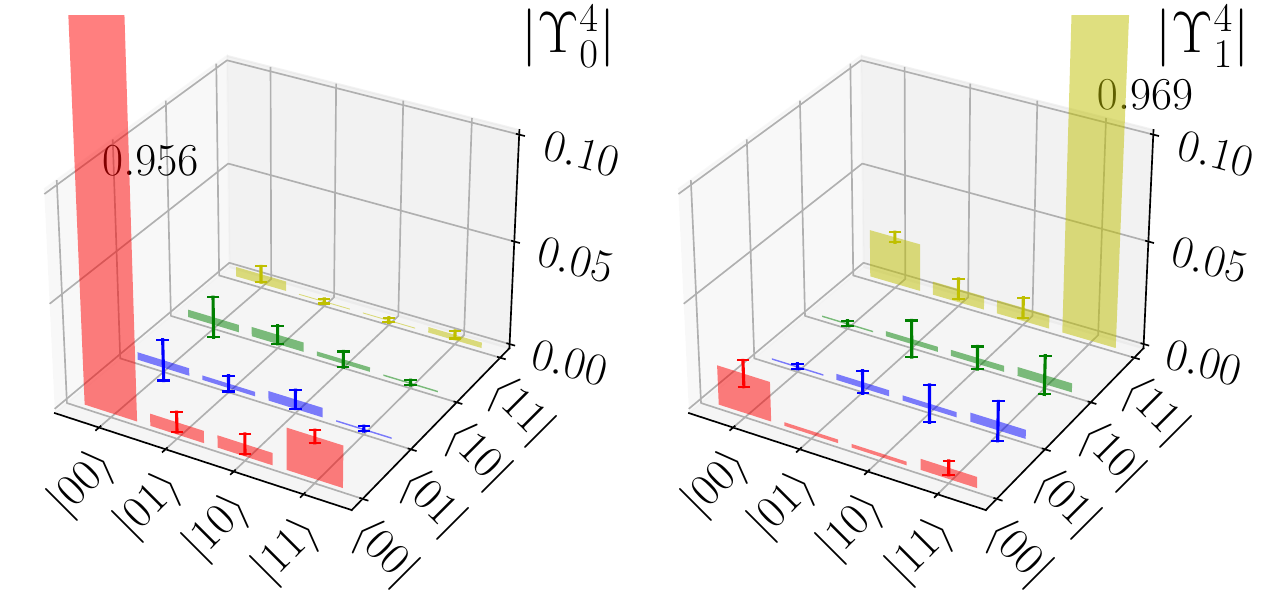}
	\caption{Absolute value of the reconstructed Choi matrices $\Upsilon_n^4$ for a measurement on qubit $\alpha=4$. Error bars are the standard deviation estimated with 5 realizations of the experiment.}
	\label{fig:q4}
\end{figure}

\begin{figure}[H]
	\centering
	\includegraphics[width=\linewidth]{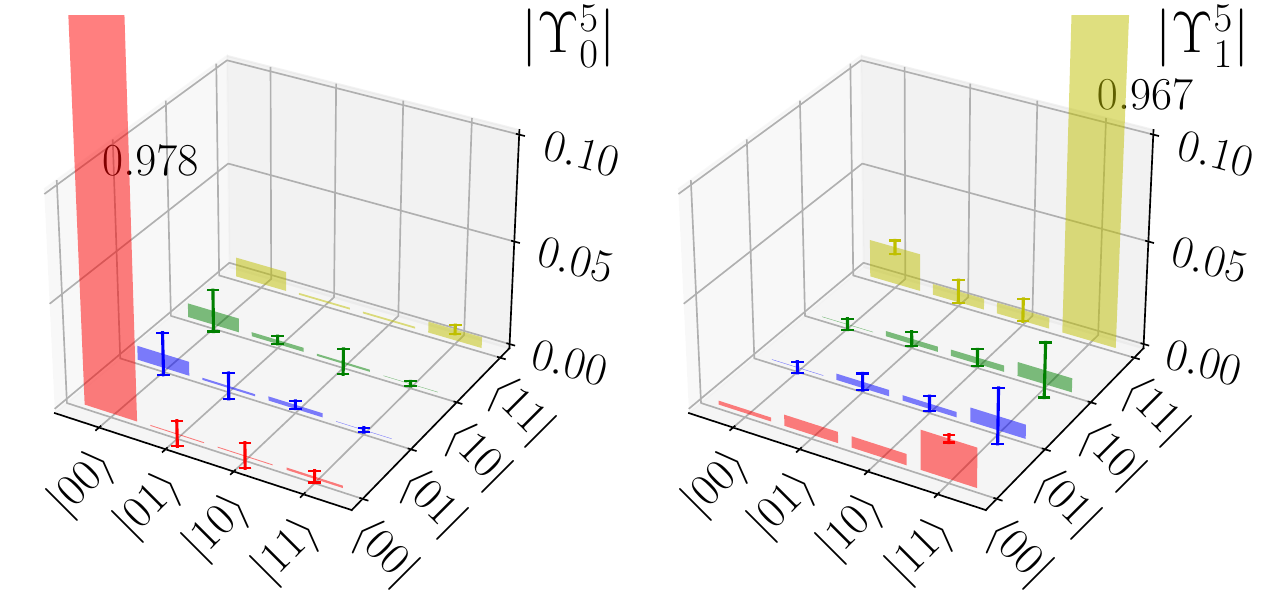}
	\caption{Absolute value of the reconstructed Choi matrices $\Upsilon_n^5$ for a measurement on qubit $\alpha=5$. Error bars are the standard deviation estimated with 5 realizations of the experiment.}
	\label{fig:q5}
\end{figure}

\begin{figure}[H]
	\centering
	\includegraphics[width=\linewidth]{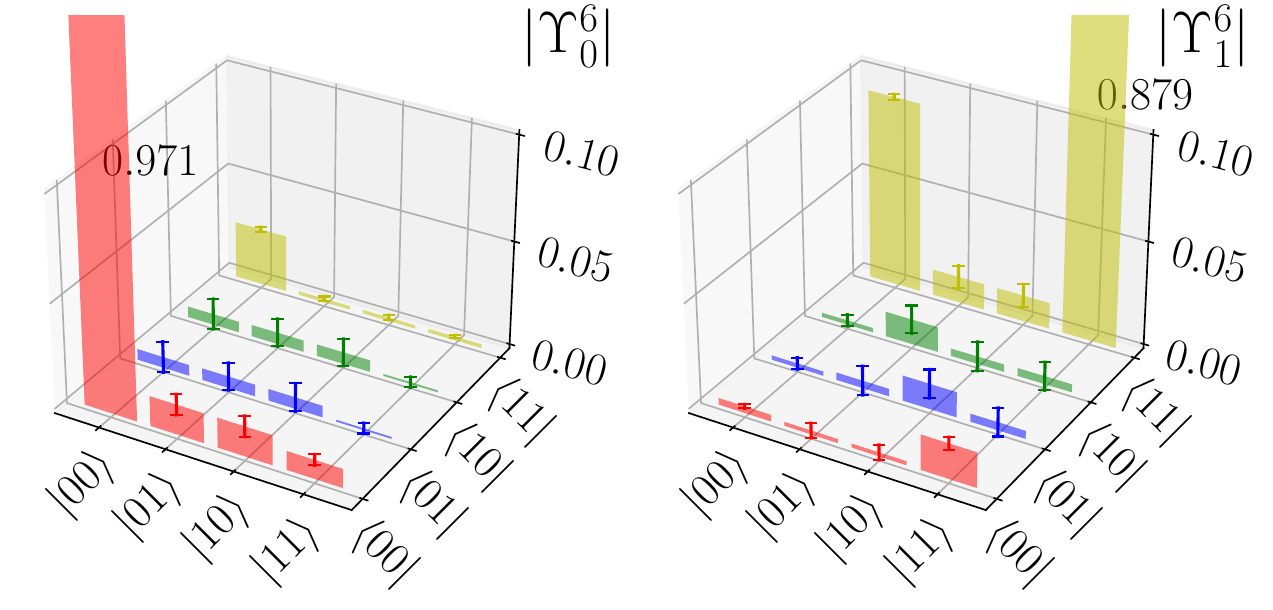}
	\caption{Absolute value of the reconstructed Choi matrices $\Upsilon_n^6$ for a measurement on qubit $\alpha=6$. Error bars are the standard deviation estimated with 5 realizations of the experiment.}
	\label{fig:q6}
\end{figure}

\subsection{II.- All experimental two-qubit Choi matrices}

In this section, we show the absolute value of all the reconstructed two-qubit Choi matrices $\Upsilon_{nm}^{\alpha\beta}$ for the four possible measurement outcomes $(n,m)\in \lbrace (0,0),(0,1), (1,0), (1,1)\rbrace$ of every pair of physically connected qubits $(\alpha,\beta)\in\lbrace (0,1),(1,2),(1,3),(3,5),(4,5),(5,6)\rbrace$ of the \emph{ibm$\_$perth} chip shown in Fig.~1(e) of the main text. All the definitions, calculation procedures, and error treatments are explained in the main text and methods. We also show the Choi matrices of the joint measurements of qubit $\alpha=3$ with respect to all other pairs in the device, which were used to calculate Fig.~4(b) of the main text.

\begin{figure}[H]
	\centering
	\includegraphics[width=\linewidth]{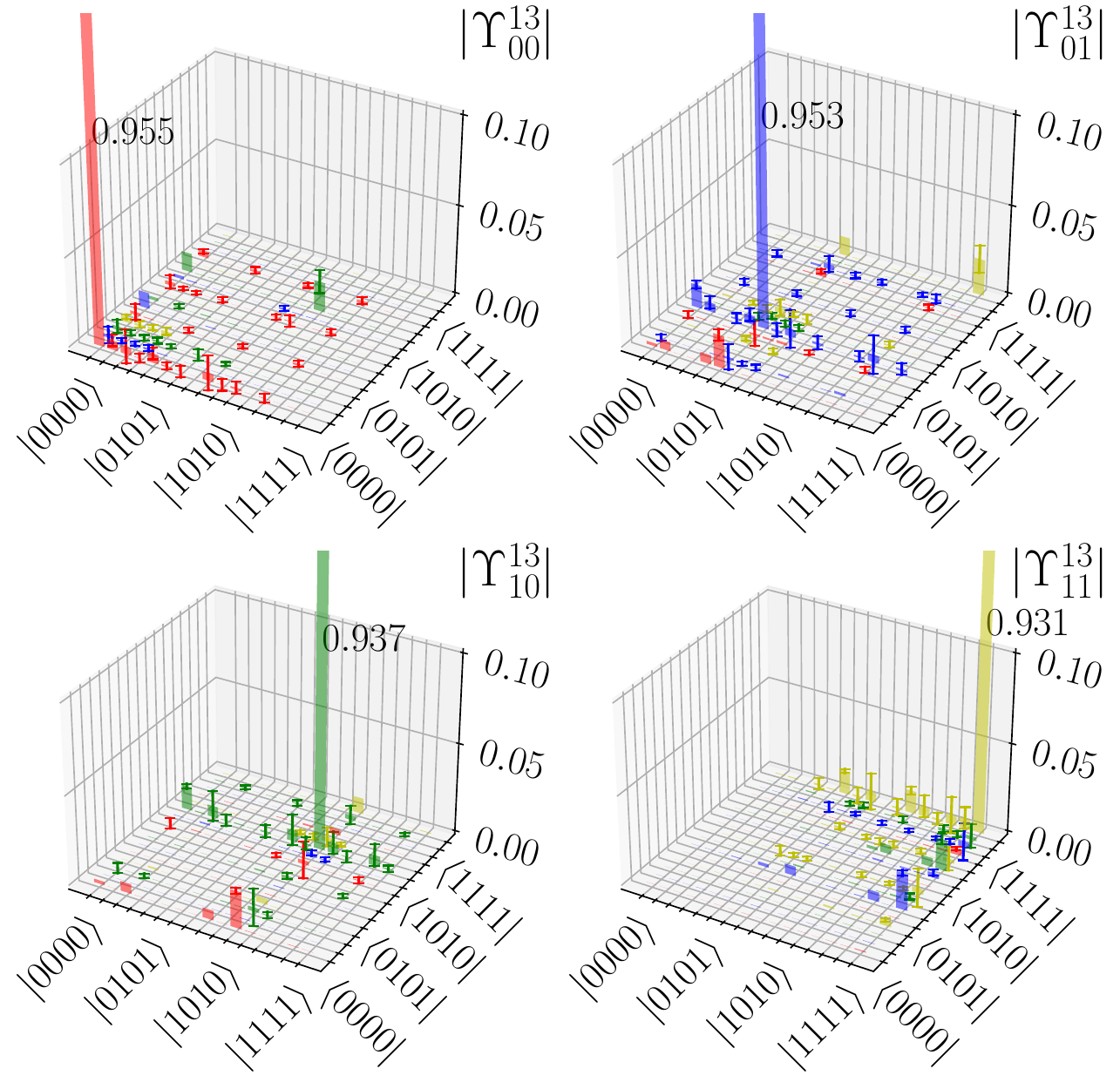}
	\caption{Absolute value of the reconstructed Choi matrices $\Upsilon_{nm}^{01}$ for a measurement on the two qubits $(\alpha,\beta)=(1,3)$. Error bars are the standard deviation estimated with 5 realizations of the experiment.}
	\label{fig:q01}
\end{figure}

\begin{figure}[H]
	\centering
	\includegraphics[width=\linewidth]{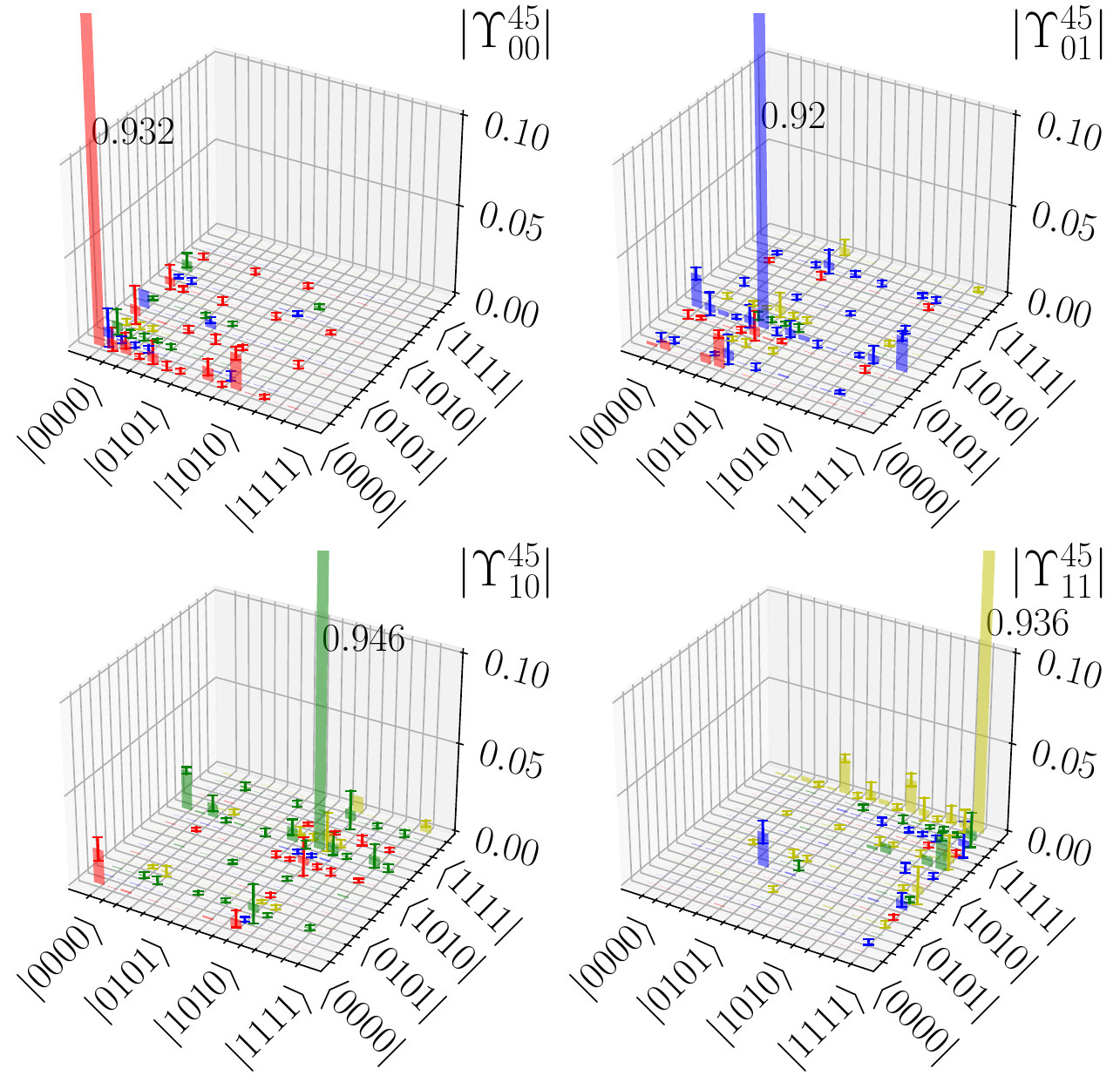}
	\caption{Absolute value of the reconstructed Choi matrices $\Upsilon_{nm}^{12}$ for a measurement on the two qubits $(\alpha,\beta)=(4,5)$. Error bars are the standard deviation estimated with 5 realizations of the experiment.}
	\label{fig:q12}
\end{figure}

\begin{figure}[H]
	\centering
	\includegraphics[width=\linewidth]{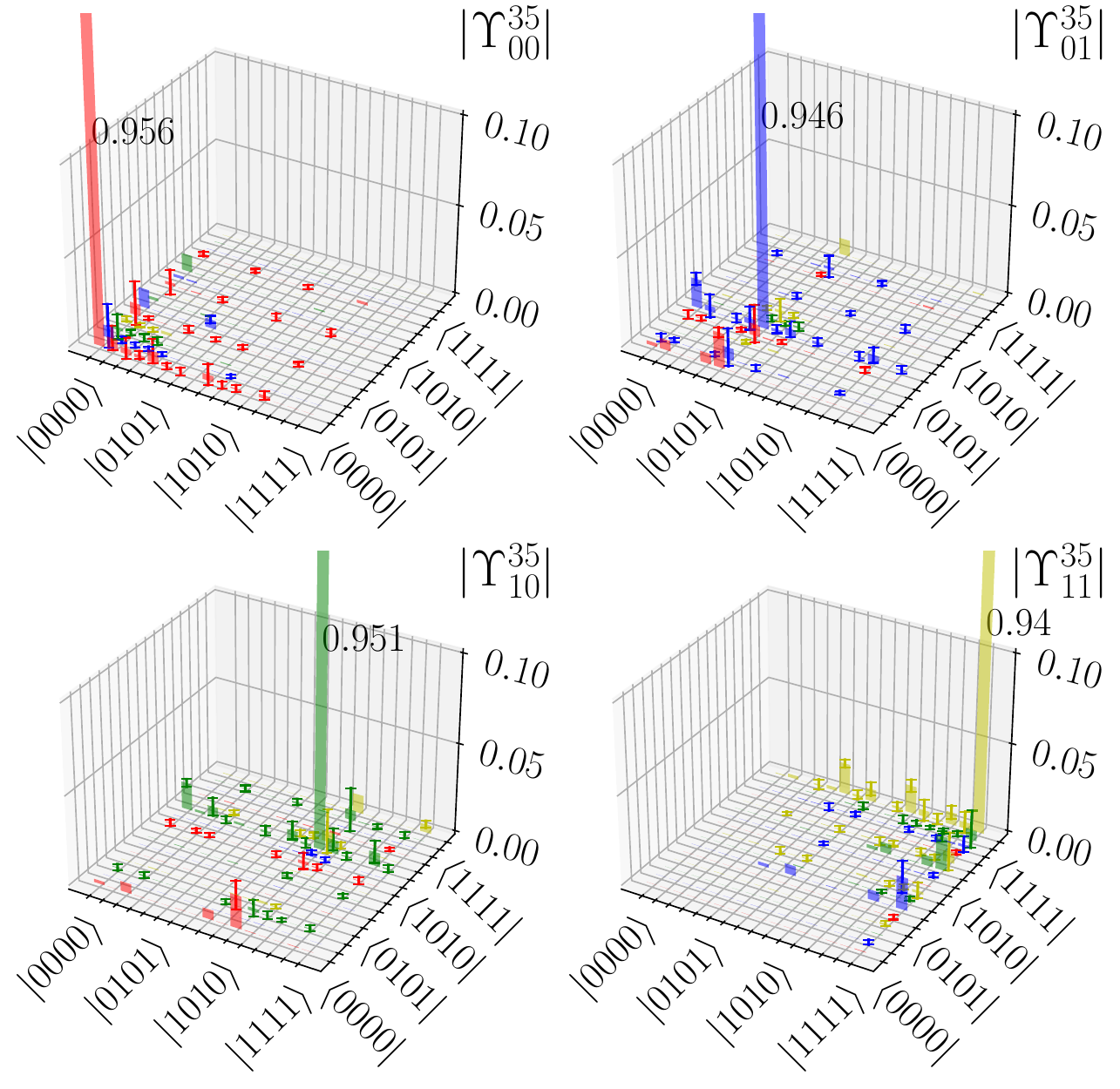}
	\caption{Absolute value of the reconstructed Choi matrices $\Upsilon_{nm}^{13}$ for a measurement on the two qubits $(\alpha,\beta)=(3,5)$. Error bars are the standard deviation estimated with 5 realizations of the experiment.}
	\label{fig:q13}
\end{figure}

\begin{figure}[h!]
	\centering
	\includegraphics[width=\linewidth]{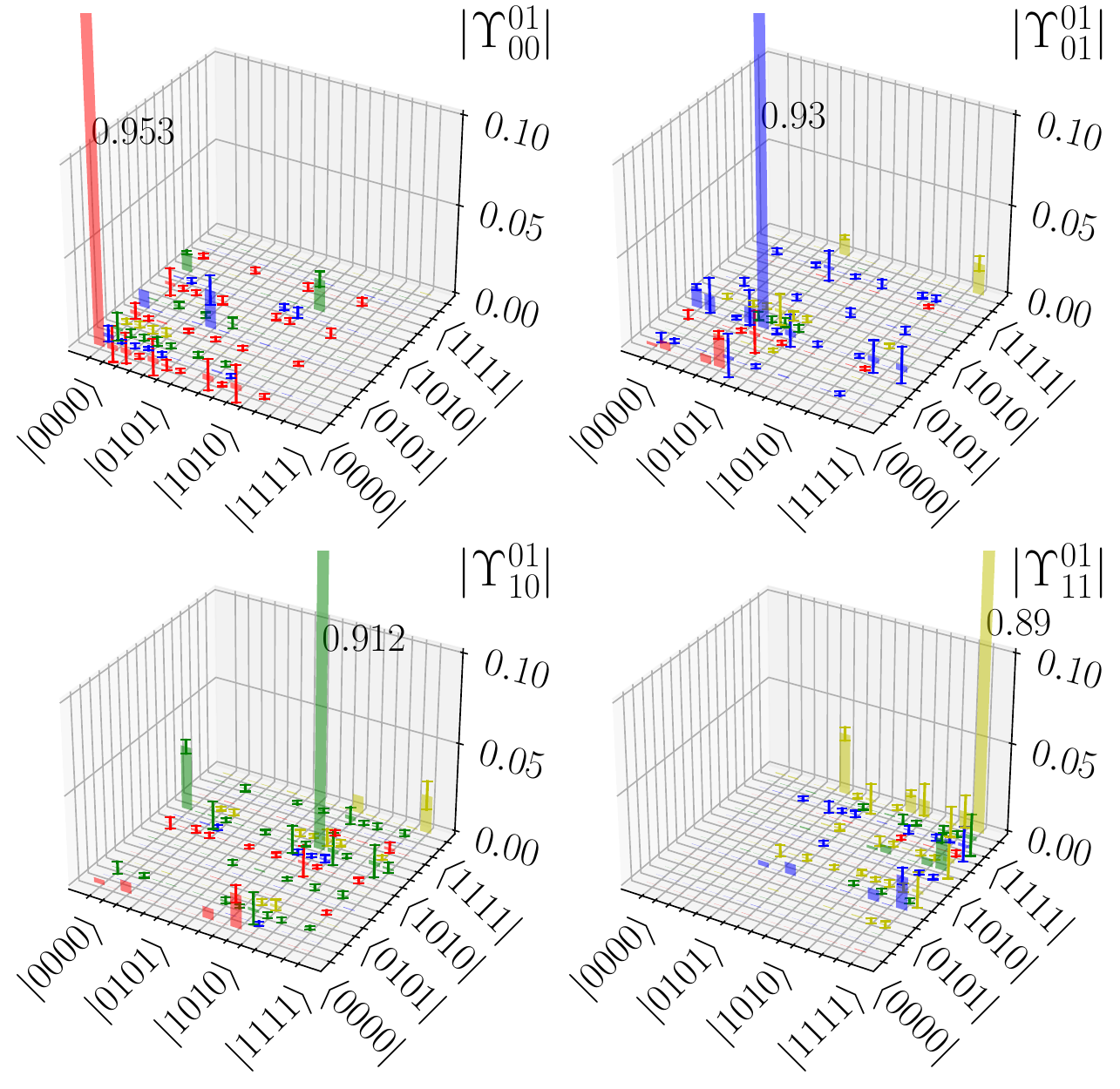}
	\caption{Absolute value of the reconstructed Choi matrices $\Upsilon_{nm}^{35}$ for a measurement on the two qubits $(\alpha,\beta)=(0,1)$. Error bars are the standard deviation estimated with 5 realizations of the experiment.}
	\label{fig:q35}
\end{figure}

\begin{figure}[H]
	\centering
	\includegraphics[width=\linewidth]{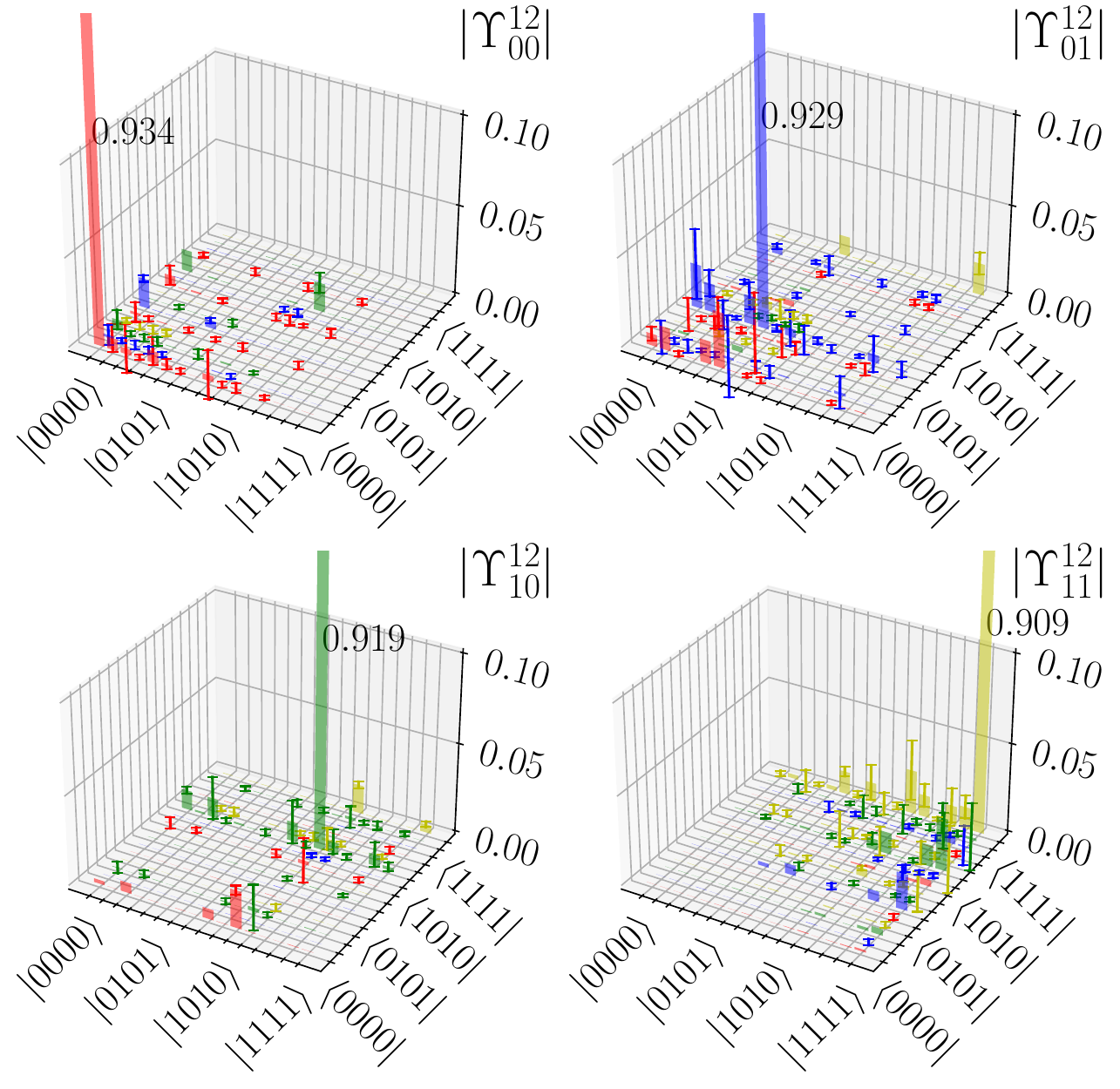}
	\caption{Absolute value of the reconstructed Choi matrices $\Upsilon_{nm}^{45}$ for a measurement on the two qubits $(\alpha,\beta)=(1,2)$. Error bars are the standard deviation estimated with 5 realizations of the experiment.}
	\label{fig:q45}
\end{figure}

\begin{figure}[H]
	\centering
	\includegraphics[width=0.99\linewidth]{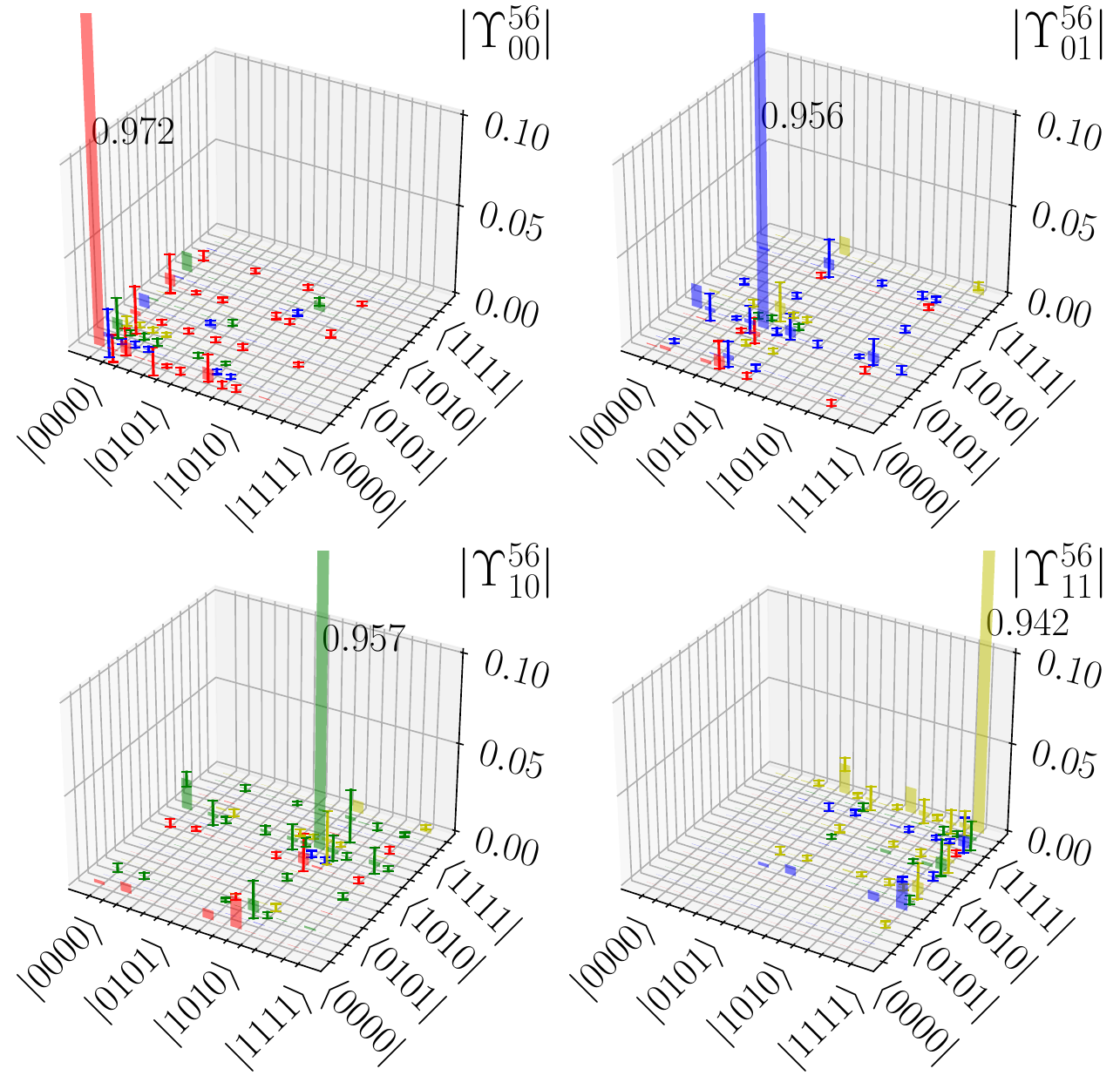}
	\caption{Absolute value of the reconstructed Choi matrices $\Upsilon_{nm}^{56}$ for a measurement on the two qubits $(\alpha,\beta)=(5,6)$. Error bars are the standard deviation estimated with 5 realizations of the experiment.}
	\label{fig:q56}
\end{figure}

\begin{figure}[h!]
	\centering
	\includegraphics[width=0.9\linewidth]{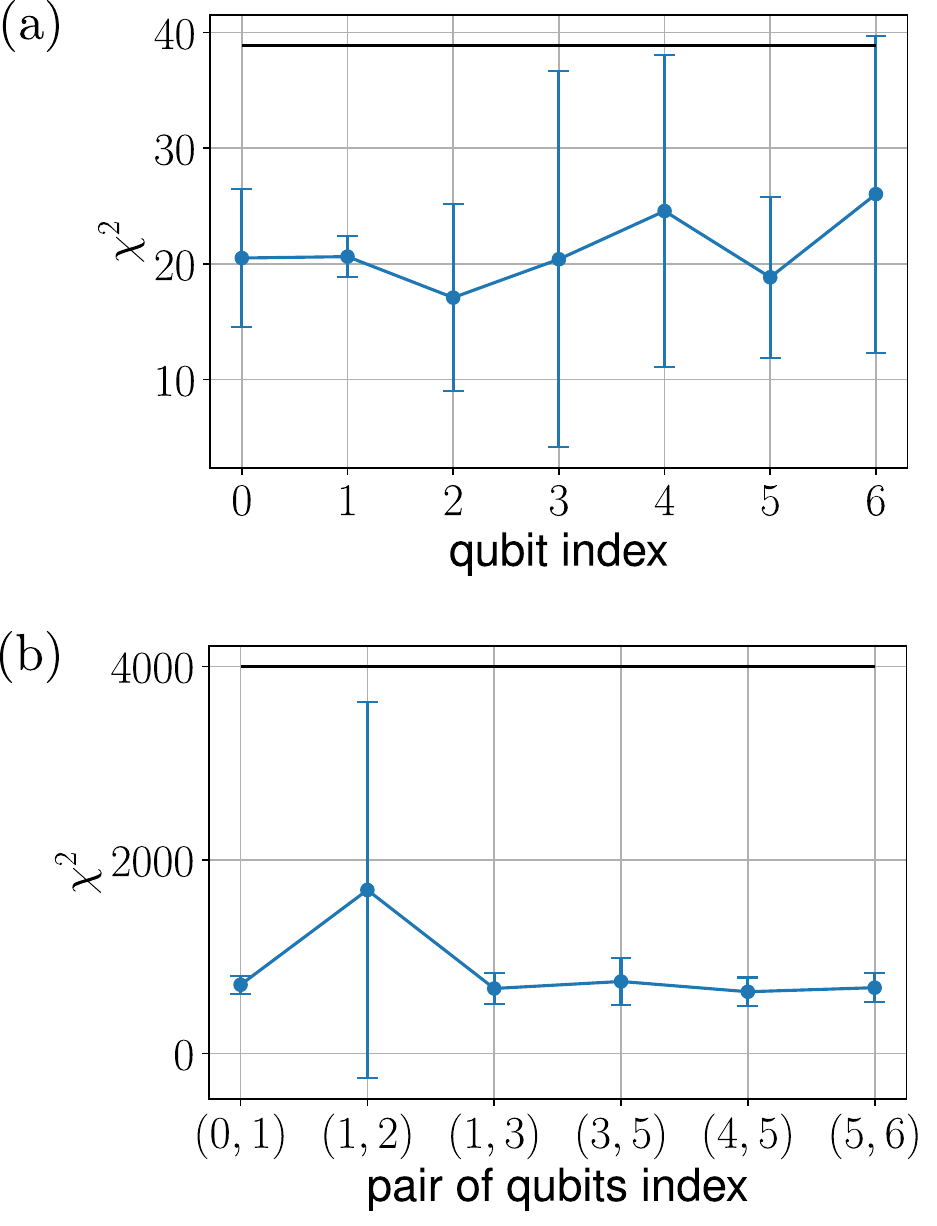}
	\caption{Goodness-of-fit of the single- and two-qubit QND-MT of \emph{ibm$\_$perth} device. All the $\chi^2$ values are within the 95\% confidence threshold indicated by the black horizontal line. Error bars are the standard deviation estimated with 5 realizations of the experiment. }
	\label{fig:gof}
\end{figure}

\section{ Supplementary methods }
\subsection{I.- Goodness-of-fit via $\chi^2$-test}

Let be $c(mn|jk)$ the counts obtained from the QND-MT of measurement process, which are used to obtain an estimator $\{ \hat\Upsilon_n \}$. The goodness-of-fit $\chi^2$ test for this data reads
\begin{enumerate}
	\item Compute the predicted probabilities,
	\begin{align}
		p(nm|jk)=\Tr[ (\hat{\mathcal{F}}_j^\dagger(\hat\Pi_m) \otimes \hat{\mathcal{F}}_k(\hat\rho)^T)\tilde\Upsilon_n],
	\end{align}
	where $\{ \hat\rho, \hat\Pi_n, \hat{\mathcal{F}}_j \}$ is the gate set estimated with GST.
	\item Compute the test statistic, 
	\begin{align}
		\chi^2 = \sum_{j=1}^M\frac{( c_i- N_{s}p_i)^2}{N_{s}p_i},
	\end{align}
	where $N_{s}$ is the number of shots used to evaluate the probabilities.
	\item Set an error probability $q$ (typically as $0.05$) and compute $\chi^2_q$, implicitly defined by
	\begin{align}
		q = \int_{\chi^2_q}^\infty P_{r}(x)dx,
	\end{align}
	where $P_{r}$ is the probability density function of a $\chi^2$ variable with mean $r$,
	\begin{align}
		P_{r}(x) = \frac{x^{(r-2)/2}e^{-x/2}}{2^{r/2}\Gamma(\frac{r}{2})}.
	\end{align}
	The mean value $r$ is given by
	\begin{align}
		r = \begin{pmatrix}
			\text{number of } \\
			\text{independent} \\
			\text{probabilities} 
		\end{pmatrix}
		-
		\begin{pmatrix}
			\text{number of} \\
			\text{parameters} \\
			\text{of the model} 
		\end{pmatrix}.
	\end{align}
	For a $N$-qubit detector, this mean value is given by
	\begin{align}
		r_N = 18^N\times( 4^{N} -1 ) - ( 2^N\times 16^N - 4^N ).
	\end{align}
	Each term from left to right corresponds to: number of circuit ($18^N$), independent probabilities per circuit ($4^{N}-1$), number of free parameters of the Choi matrices ($2^N\times16^N$), and completeness constraint ($4^N$). For the cases of single- and two- qubits detectors, these are $r_1=26$ and $r_2=3852$, respectively. 
	
	\item Reject the hypothesis if $\chi^2\geq\chi^2_\alpha$.
\end{enumerate}

We apply this procedure to quantify the goodness-of-fit of our characterization of the \emph{ibm$\_$perth} device with QND-MT presented. As shown in Suplementary Figure~\ref{fig:gof}, both single- (a) and two-qubit (b) characterizations have a $\chi^2$ value below the threshold $\chi_q^2$ (black horizontal line) and therefore are consistent with the experimental data with confidence $95\%$ ($q=0.05$).

\end{document}